\documentclass[letterpaper]{JHEP3}

\usepackage{graphicx}
\usepackage{amsmath,amsthm, amssymb}

\newcommand{\eq}{\begin{equation}}
\newcommand{\eqe}{\end{equation}}

\newcommand{\eqa}{\begin{eqnarray}}
\newcommand{\eqae}{\end{eqnarray}}

\newcommand{\m}{\mu}
\newcommand{\n}{\nu}
\newcommand{\e}{\epsilon}

\title{\bf Supertwistor space for 6D maximal super Yang-Mills  }
\author{
Tristan Dennen\footnote{Email: tdennen@physics.ucla.edu}$^{~1}$, Yu-tin Huang\footnote{Email: yhuang@physics.ucla.edu}$^{~1}$, and
Warren Siegel \footnote{Email: siegel@insti.physics.sunysb.edu}$^{~2}$
\\ \\ \\
\it $^1$ Department of Physics and Astronomy,\\ UCLA,\\
Los Angeles, CA 90095-1547, USA
\\
\\
\it $^2$ C.N. Yang Institute for Theoretical Physics,\\
\it Stony Brook University, \\
\it Stony Brook, NY 11794-3840, USA}
\abstract{
6 dimensional maximal super Yang-Mills on-shell amplitudes are formulated in superspace using 6 dimensional spinors. The 3,4,5-point tree amplitudes are obtained by supersymmetrizing their bosonic counterparts and confirmed through the BCFW construction. In contrast to 4 dimensions this superspace is non-chiral, reflecting the fact that one cannot differentiate MHV from $\overline{{\rm MHV}}$ in 6 dimensions. Combined with unitarity methods, this superspace should be useful for the study of multi-loop D dimensional maximal super Yang-Mills and gravity amplitudes. Furthermore, the non-chiral nature gives a natural framework for an off-shell construction. We show this by matching our result with off-shell D=4 N=4 super Yang-Mills amplitudes, expressed in projective superspace.    
}
\preprint{YITP-SB-09-29\\
UCLA/09/TEP/09/86} 

\keywords{superspace, maximal super Yang-Mills, BCFW}

\begin{document}

\numberwithin{equation}{section}


\section{Introduction} 
In recent years many surprising results were discovered in the S-matrix of maximal supersymmetric theories in 4 dimensions. These include new symmetries and structures \cite{Drummond:2008vq}, representations \cite{Witten:2003nn,Cachazo:2004kj,Britto:2004ap,Bern:2008qj,Mason:2009sa, ArkaniHamed:2009dn} of tree-level amplitudes, and unexpected UV behaviour in loop perturbation theory \cite{Bern:2006kd,Berkovits:2009aw,new,Bossard:2009sy}. Many of these advancements rely heavily on newly  developed on-shell methods such as recursion relations to construct tree amplitudes, and generalized unitarity to obtain loop corrections by simply sewing tree amplitudes. More precisely, one can now use either the CSW method \cite{Cachazo:2004kj}, which constructs general amplitudes from MHV vertices, or the BCFW \cite{Britto:2004ap} construction, which expresses an $n$-point amplitude as direct products of lower point amplitudes, to efficiently construct tree amplitudes for either gauge or gravity theory. Modern unitarity methods \cite{Britto:2004nc} then allow one to construct loop amplitudes by expressing them in terms of a set of integrals that reproduces the cuts of the amplitude. Tree amplitudes are then used to construct the coefficients of these integrals.

The major unsatisfactory aspects in these current approaches is their reliance on 4 dimensional spinor-helicity formalism \cite{helicity, Ferber:1977qx}, while many interesting questions are inherently D dimensional. For example in the study of divergences in maximal supersymmetric theories, one usually encounters various bounds (at given loop level) on the dimension at which the first potential divergence should appear \cite{Bern:2006kd, Bern:1998ug}. To study this bound one is required to compute the divergences of the D dimensional theory. On the other hand even in QCD one loop amplitudes, D dimensional tree amplitudes are useful for obtaining rational terms when using unitarity methods \cite{qcd}. Therefore a spinor helicity formalism similar to 4 dimensions will be helpful for these purposes.   

Since physical degrees of freedom are completely determined by its super Poincar\'e quantum numbers, the power of spinor helicity formalism is then to represent these quantum numbers covariantly using unconstrained variables. There has been recent progress in constructing general D$\neq$4 spinor helicity formalism \cite{Boels:2009bv} and for D=10 \cite{Berkovits:2009by}, though the variables are constrained. Here we focus on 6 dimensions where the spinor-helicity formalism is very similar to 4 dimensions, as recently demonstrated by Cheung and O'Connell \cite{Cheung:2009dc}. The idea is to start in 6 dimensions where the Lorentz group SO(5,1) has the covering group SU$^*$(4). The vector forms an antisymmetric representation of SU$^{*}$(4), and the on-shell condition is naturally solved by introducing SU$^{*}$(4) spinors, $P^{AB}=\lambda^{Aa}\lambda^{B}_a,\,P_{AB}=\tilde{\lambda}_{A\dot{a}}\tilde{\lambda}^{\dot{a}}_{B}$. The indices $a,\dot{a}$ transform under the 6 dimensions little group SO(4)$\simeq$SU(2)$\times$ SU(2). In fact, these spinors can be viewed as (half)part of the spinor representation of the six dimensional conformal group SO$^{*}$(8), i.e. they are twistors\cite{Penrose:1986ca}. In this light, their property of being solutions to the massless constraint follows directly from twistors being solutions to conformal constraints.

In this paper we will introduce Grassmann variables along with the spinors to form an on-shell superspace. As we will demonstrate, the Yang-Mills field strength is in the $(\frac{1}{2},\frac{1}{2})$ representation of the little group. Since for maximal N=(1,1) theory the full multiplet should be contained in a single superfield, the non-chiral nature of the field strength then implies a non-chiral on-shell superspace. These Grassmann variables are the fermionic pieces of the spinor representation(supertwistors) of the six dimensional superconformal group OSp$^{*}$(8$|$2N). Since the supertwistors are self-conjugate, we covariantly truncate these variables using the SU(2)$\times$SU(2) R indices. Compared to 4 dimensions whose twistor is not self-conjugate, one simply takes the chiral twistor at the loss of manifest parity symmetry, here it is replaced by the loss of R symmetry. Being non-chiral has the advantage of representing the amplitudes in a more symmetric fashion, instead of viewing the amplitudes from the MHV (or $\overline{{\rm MHV}}$) point of view.\footnote{More precisely, from the view point of self dual (or anti-self dual) super Yang-Mills, which is naturally expressed in terms of chiral superspace \cite{Chalmers:1996rq}.} 

This unification of MHV and $\overline{{\rm MHV}}$ amplitudes in this superspace hints at 4 dimensional off-shell superspace which must be non-chiral. In fact the splitting of R-indices, when reduced to 4 dimensions, is similar to the 4 dimension projective superspace described in \cite{Hatsuda:2007wr}. We will see that our 4-point amplitude using 6 dimensions spinors shares the same form with that recently derived in 4 dimensions projective superspace \cite{Hatsuda:2008pm}. The fact that one may understand 4 dimensions off-shell superspace from 6 dimensions on-shell is similar to the usual story of viewing the conformal group SO(2,4) as the Lorentz group in 6 dimensions acting on the projective (modding out the scale)  light-cone ($p^2=0$). The theory on the light-cone is four dimensional and off-shell.

We begin with a discussion of 6 dimensional spinors similar to Cheung and O'Connell. In section 3 we introduce Grassmann variables in the spirit of Ferber \cite{Ferber:1977qx} and construct the N=2 superspace. In section 4 we obtain the super amplitudes by simply supersymmetrizing the 3, 4 and 5-point amplitudes derived in \cite{Cheung:2009dc}. In section 5 we rederive the previous result using BCFW. Finally we show the application of this approach to loop amplitudes by reproducing the one-loop four-point structure of maximal SYM in D dimensions \cite{Bern:1997nh}.

\section{6 dimensional on-shell spinors}
We review the 6 dimensional spinor-helicity formalism recently developed by Cheung and O'Connell \cite{Cheung:2009dc}. We will present it in parallel with the familiar 4 dimensional results. In 6 dimensions Minkowski space the Lorentz group is SO(5,1) whose covering group is  SU$^*$(4). The vector is in the anti-symmetric representation of SU$^*$(4), and the scalar product of two vectors is defined as a contraction with the SU$^{*}$(4) invariant tensor $\e_{ABCD}$. For simplicity we drop the $^*$ from now on. For a null vector one has   
\eq
6\,D: p^\mu=p^{AB},\;p^2=0\rightarrow \epsilon_{ABCD}p^{AB}p^{CD}=0\rightarrow p^{AB}=\lambda^{A a}\lambda^B\,_{a},
\eqe
where the spinors $\lambda^A_a$ are pseudo real, $A$ is the SU(4) index and $a,\dot{a}$ are the SU(2) indices.\footnote{One can work in other signatures, in the Wick rotated SO(3,3) the covering group would be SL(4), $a,\dot{a}$ transform under SL(2) and the spinors are real.} The bi-spinor form of the momentum solves the on-shell constraint since there are no 4 component totally anti-symmetric tensors in SU(2). One can also represent the momentum in the anti-fundamental representation:  
\eq
p_{AB}=\frac{1}{2}\epsilon_{ABCD}p^{CD}=\tilde{\lambda}_{A\dot{b}}\tilde{\lambda}_{B}^{\dot{b}},\;\; \lambda^A_a\tilde{\lambda}_{A\dot{a}}=0.
\eqe
One can also understand this solution by counting components. A null vector in 6 dimensions has 5 components including a scale factor, meanwhile $\lambda^{Aa}$ has $4\times2=8$ components and the SU(2) invariance removes 3 of them. Since the definition of little group is the transformation group that leaves the null momentum invariant, the SU(2) indices on the spinors correspond to the 6 dimensions little group SO(4), whose covering group is SU(2)$\times$SU(2).

This is similar to 4 dimensions on-shell momentum which has 3 components. We write the 4 dimensional real momentum in terms of spinors
\eq
4\,D:p_{\alpha\dot{\alpha}}=\lambda_{\alpha}\tilde{\lambda}_{\dot{\alpha}}.
\eqe
With $\lambda_\alpha$ being complex and $\tilde{\lambda}_{\dot{\alpha}}=\pm\bar{\lambda}_{\dot{\alpha}}$ in Minkowski space, one also has $4-1=3$ components, where the 1 is from the invariance of $p_{\alpha\dot{\alpha}}$ under U(1) rotation $\lambda_\alpha\rightarrow e^{i\theta}\lambda_\alpha,\;\tilde{\lambda}_{\dot{\alpha}}\rightarrow e^{-i\theta}\tilde{\lambda}_{\dot{\alpha}}$.

Note that in arbitrary dimensions, one can always represent an on-shell momentum in bi-spinor form by first finding the solutions to the Dirac equation. These solutions can then be used to construct the null momenta. However since the solution of the Dirac equation is non-covariant, this approach will be less useful for analytic analysis of the amplitudes. An important distinction for the above (4)6 dimensional discussion, is that the spinors ``automatically" satisfy the Dirac equation, they are unconstrained variables. This matching between massless degrees of freedom and the moding of the little group from the spinors only exists in 3,4 and 6 dimensions.  A demonstration of this difficulty can be seen in the recent 10 dimensional twistor construction \`{a} la Berkovits\cite{Berkovits:2009by}. There the 10 dimensional null vector is constructed using a pure spinor $\lambda$ and a Weyl spinor $\pi$, $p^\mu=\lambda\gamma^\mu\pi$. There is also a gauge invariance $\delta\pi=(\gamma^{\mu\nu}\lambda)\Omega_{\mu\nu}$ which gives the correct counting for an on-shell momentum, 22+32-45=9\footnote{There are 22 degrees of freedom for a pure spinor.}. However the gauge group is SO(10) which is larger than the little group SO(8), which results in residual gauge invariance in the components of the supertwistor field.

These 6 dimensional spinors are half of the 6 dimensional twistor. This twistor is in the irreducible spinor representation of the 6 dimensional conformal group SO(6,2)=SO(8)*, an 8 dimensional chiral spinor which is composed of a 6 dimensional chiral and anti-chiral spinor\cite{Penrose:1986ca}. The $\lambda^{Aa}$ and $\tilde{\lambda}_{A\dot{a}}$ are part of the chiral and anti-chiral twistor, which is equivalent representations due to triality, respectively. Since it is pseudoreal, the twistor and it's equivalent complex conjugate form a SU(2) doublet. This SU(2) then becomes the SU(2) little group for the 6 dimensional spinors. Even though in this paper we are interested in the N=(1,1) theory which is non-conformal, it is still useful to analyse scattering amplitudes of non-conformal massless theories in twistor space. A well known example is the study of Yang-Mills amplitudes using 4 dimensional twistors. Note that one anticipates this understanding of 6 dimensional spinor helicity in terms of twistors will play an important role for the analysis of the (2,0) theory which is expected to be superconformal.  

Lorentz invariants are constructed by contracting the SU(4) indices:
\eqa
\nonumber\e^{ABCD}(\tilde{\lambda}_1)_{A\dot{a}}(\tilde{\lambda}_2)_{B\dot{b}}(\tilde{\lambda}_3)_{C\dot{c}}(\tilde{\lambda}_4)_{D\dot{d}}&=&[1_{\dot{a}}2_{\dot{b}}3_{\dot{c}}4_{\dot{d}}]\\
\nonumber \e_{ABCD}(\lambda_1)^{A}\,_{a}(\lambda_2)^{B}\,_{b}(\lambda_3)^{C}\,_{c}(\lambda_4)^{D}\,_{d}&=&\langle1_{a}2_{b}3_{c}4_{d}\rangle,\\
\nonumber (\lambda_i)^A_{a}(\tilde{\lambda}_j)_{A\dot{a}}=\langle i_a|j_{\dot{a}}]\rightarrow \det(\langle i_a|j_{\dot{a}}])&=&-2p_i\cdot p_j\\
\nonumber \langle \lambda_a|\displaystyle{\not}p_1\displaystyle{\not}p_2\displaystyle{\not}p_3|\lambda_b\rangle&=&\lambda^A_a (p_1)_{AB}(p_2)^{BC}(p_3)_{CD}\lambda^D_b\\
\nonumber[ \tilde{\lambda}_{\dot{a}}|\displaystyle{\not}p_1\displaystyle{\not}p_2\displaystyle{\not}p_3\displaystyle{\not}p_4|\lambda_b\rangle&=&\tilde{\lambda}_{A\dot{a}} (p_1)^{AB}(p_2)_{BC}(p_3)^{CD}(p_4)_{DE}\lambda^E_b.
\eqae
Note that a chiral and an anti-chiral spinor can only be contracted with an even number of momenta. These spinors can be expressed in terms of momenta in a non-covariant way. Furthermore, when the momenta are restricted to a 4 dimensions subspace, all of the above Lorentz invariants can be rewritten in terms of 4 dimensional spinors. We demonstrate these properties in appendix (\ref{4D}). In light of the proliferation of indices, we make a brief list:
\begin{itemize}
  \item $A,B,C\cdot\cdot$ are SU(4) indices of the 6 dimensional Lorentz group
  \item $a,b,\cdot\cdot,\dot{a},\dot{b},\cdot\cdot,$ are the SU(2)$\times$SU(2) little group indices 
  \item $I,J,K,\cdot\cdot$ are the R-symmetry indices
  \item $i,j,k,\cdot\cdot$ labels the external line
  \item $\mu,\nu,\cdot\cdot$ are the spacetime index in any dimensions 
  \item $\alpha,\beta,\cdot\cdot\dot{\alpha},\dot{\beta},\cdot\cdot$ are 4D SL(2,C) indices 
\end{itemize}

In 4 dimensions the polarization vectors are written as
\eq
4\,D: (\epsilon^\mu_+)_{\beta\dot{\beta}}=\frac{\tilde{\lambda}_{\dot{\beta}}\kappa_\beta}{\lambda^\alpha\kappa_\alpha}=\frac{|\tilde{\lambda}]\langle\kappa|}{\langle\lambda\kappa\rangle},\;\;(\epsilon^\mu_-)_{\beta\dot{\beta}}=\frac{\lambda_\beta\tilde{\kappa}_{\dot{\beta}}}{\tilde{\lambda}^{\dot{\alpha}}\tilde{\kappa}_{\dot{\alpha}}}=\frac{|\lambda\rangle[\tilde{\kappa}|}{[\tilde{\lambda}|\tilde{\kappa}]},
\label{polar4}
\eqe  
where $\kappa$ is the spinor for an arbitrary null vector $k_{\alpha\dot{\alpha}}$ with $k\cdot p\neq0$. Similarly in 6 dimensions\footnote{The object $\frac{1}{[i_{\dot{a}}|j_b\rangle}$ is defined as the inverse matrix $([i_{\dot{a}}|j_b\rangle)^{-1}=\frac{[i^{\dot{a}}|j^b\rangle}{s_{ij}}$.}
\eqa
\label{polar}
\nonumber6\,D: (\epsilon^{\mu}_{a\dot{a}})^{AB}&\equiv&\sqrt{2}\frac{\lambda^{[A}\,_{a}\kappa^{B]}\,_{c}}{\kappa^{D}\,_{c}\tilde{\lambda}_{D}\,^{\dot{a}}}=\sqrt{2}\frac{|^{[A}\lambda_a\rangle\langle \kappa_c\,^{B]}|}{\langle \kappa_c|\lambda^{\dot{a}}]}=\sqrt{2}\frac{|^{[A}\lambda_a\rangle(\displaystyle{\not}k|\lambda_{\dot{a}}])}{det\langle \kappa|\lambda]}^{B]}\\
\;(\epsilon_{\mu a\dot{a}})_{AB}&\equiv&\sqrt{2}\frac{\tilde{\lambda}_{A\dot{a}}\tilde{\kappa}_B\,^{\dot{c}}}{\lambda^D\,^{a}\tilde{\kappa}_D\,^{\dot{c}}}=\sqrt{2}\frac{|_{[A}\tilde{\lambda}_{\dot{a}}][\tilde{\kappa}_{\dot{c}B]}|}{[\tilde{\kappa}_{\dot{c}}|\lambda^a\rangle}=\sqrt{2}\frac{|_{[A}\tilde{\lambda}_{\dot{a}}](\,\displaystyle{\not}k|\lambda_a\rangle)_{B]}}{det[\tilde{\kappa}|\lambda\rangle}. 
\eqae
Again $\kappa^{A\alpha}$ is the spinor for some reference null momenta $k^{AB}$, and $\epsilon^\mu\,_{a\dot{a}}(\epsilon_{\mu})_{b\dot{b}}=C_{ab}C_{\dot{a}\dot{b}}$, where $C_{ab}=C_{\dot{a}\dot{b}}=\left(\begin{array}{cc}0 & -i \\i & 0\end{array}\right)$. The determinant in the denominator occurs over the little group indices. Note that the polarization vectors transform in the ($\frac{1}{2}$,$\frac{1}{2}$) representation of the little group. In both cases, one can easily show the polarization vectors satisfy $p^\mu\e_\mu=0$, and an arbitrary redefinition of the reference spinor translates into a gauge transformation.

The field strength $F_{\mu\nu}=p_\m\e_\n-p_\n\e_\m$ also has a simple expression in terms of spinors. In 4 dimensions, using the definition of $\e$ in (\ref{polar4}) the field strength naturally separates into a chiral and anti-chiral piece:
$$F_{\mu\nu}=F_{\alpha\dot{\alpha}\beta\dot{\beta}}=C_{\dot{\alpha}\dot{\beta}}f_{\alpha\beta}+C_{\alpha\beta}\tilde{f}_{\dot{\alpha}\dot{\beta}}\rightarrow f_{\alpha\beta}=\lambda_\alpha\lambda_\beta,\;\tilde{f}_{\dot{\alpha}\dot{\beta}}=\tilde{\lambda}_{\dot{\alpha}}\tilde{\lambda}_{\dot{\beta}}.$$
Using (\ref{polar}) for 6 dimensions one obtains
\eq
F_{\m\n}=(F_{AB,CD})_{a\dot{a}}=-3(\e_{ABCE}\lambda^E\,_{a}\tilde{\lambda}_{D\dot{a}}+\e_{DBCE}\lambda^E\,_{a}\tilde{\lambda}_{A\dot{a}}-\e_{ABDE}\lambda^E\,_{a}\tilde{\lambda}_{C\dot{a}}-\e_{DACE}\lambda^E\,_{a}\tilde{\lambda}_{B\dot{a}}).
\eqe
One can contract the field strength with the SU(4) invariant tensor $\e$ to obtain the following quantities 
\eqa
\nonumber (F^{AB}\,_{CD})_{a\dot{a}}=\frac{1}{2}\e^{ABEF}F_{EFCD}=\frac{-3}{2}\lambda^{[A}\,_a\tilde{\lambda}_{[C\dot{a}}\delta^{B]}_{D]},\\
(F^E\,_D)_{a\dot{a}}=\frac{1}{3!}\e^{EABC}(F_{ABCD})_{a\dot{a}}=\lambda^E\,_{a}\tilde{\lambda}_{D\dot{a}}.
\label{def}
\eqae 
The last expression will be the one that appears naturally in amplitudes.

\section{6 dimensions N=(1,1) superspace}
Since both the polarization vector and the field strength appear as  ($\frac{1}{2}$,$\frac{1}{2}$) tensors in SU(2)$\times$ SU(2), this implies that the on-shell superspace must be non-chiral as well. Note that the chiral N=(1,0) super Yang-Mills and the mysterious N=(2,0) theory use chiral on-shell superspaces; however the former is not maximal\footnote{For non-maximal theories, the on-shell states cannot be contained in a single on-shell superfield. } and the later does not contain a vector gauge field. 

Recent constructions of the S-matrix for maximal gauge and gravity theories make use of  4 dimensional  supertwistor space. Here we construct the 6 dimensional N=2 on-shell superspace in similar fashion, i.e. by introducing Grassmann variables $\eta^I_a$\footnote{These $\eta$ variables appear in a similar fashion as with the 4 dimensional N=4 theory. We refer to \cite{ArkaniHamed:2008gz} for a detailed discussion of its properties.}, where $I$ is the R index and $a$ is the little group index, one can arrive at the usual superspace by contracting the little group indices with the spinors. In 4 dimensions $I=1,2,3,4$ and the little group is U(1), under which the Grassmann variables transform as $\eta^I\rightarrow e^{-i\theta}\eta^I,\bar{\eta}_I\rightarrow e^{i\theta}\bar{\eta}_I$. The relation to the usual superspace can be seen with the help of the spinors 
$$4\,D:\;\theta^{I\alpha}=\lambda^\alpha \eta^I,\;\bar{\theta}_{J\dot{\alpha}}=\tilde{\lambda}_{\dot{\alpha}} \bar{\eta}_J.$$ 
Note that in a sense one contracts with respect to the little group.

One can do similar for 6 dimensions. Maximal super Yang-Mills in 6 dimensions has N=(1,1) supersymmetry with R-symmetry group USp(2)$\times$USp(2)=SU(2)$\times$SU(2). We introduce $\eta^{a I}$ and $\tilde{\eta}_{\dot{a}I'}$ where the $I,I'$ are the SU(2)$_R$ symmetry indices. Note that $\eta$ and $\tilde{\eta}$ are complex and independent. They are the fermionic part of the chiral and anti-chiral supertwistor; the spinor representation of OSp$^*$(8$|$2). The full 6 dimension superspace variables are then 
$$6\,D:\;q^{AI}=\lambda^A_a \eta^{a I},\;\tilde{q}_{AI'}=\tilde{\lambda}_{A\dot{a}} \tilde{\eta}^{\dot{a}}_{I'}.$$

In 4 dimensions maximal super Yang-Mills (as well as gravity), one can express the full amplitude using either chiral or anti-chiral superspace, i.e. only half of the full superspace, since this is enough to contain all physical degrees of freedom. This is due to the self-CPT conjugate nature of the physical spectrum. In 6 dimensions we have similar result. However since the supertwistors are self-conjugate, only half of the degrees of freedoms for $\eta^{a I}$ and $\tilde{\eta}_{\dot{a}I'}$are independent. Therefore to construct our on-shell superspace we need to truncate the $\eta,\tilde{\eta}$s. Since we wish to use the little group index to label our states, we will truncate using the R-indices.

Note that this situation is equivalent to the issue of trying to construct off-shell N$>1$ superspace, where chiral constraints usually lead to field equations. One of the well known examples is the N=2 harmonic superspace \cite{Galperin:1984av} in 4 dimensions. Here one introduces harmonic variables $u^{\pm}_{I}$ to parameterize the SU(2)/U(1) coset. These variables are then used to separate the $\theta$ variables into two separate sets ($\theta^{+}_\alpha=u^{+}_I\theta^I_\alpha,\bar{\theta}^{+}_{\dot{\alpha}}=u^{+I}\bar{\theta}_{I\dot{\alpha}}$) and $+\rightarrow -$. Then the prepotential, which contains the physical gauge field, depends only on a subspace (the ``analytic superspace") which only includes $\theta^+,\bar{\theta}^+$. The harmonic variables can be viewed as providing a linear combination of the R-symmetry index, and therefore separating the supercharges into subsets.\footnote{Of course these new bosonic R-coordinates also provide the infinite auxiliary fields that are necessary to close the susy algebra off-shell. Different choices (or a subset) of these coordinates represent different off-shell formulations, for example there is also the N=2 projective superspace \cite{Karlhede:1984vr}.}

Therefore we can use the two harmonics coming from our SU(2)$\times$SU(2) R symmetry to construct our half superspace, i.e. we choose our subspace to include only $q^{A+}=u^+_Iq^{AI},\tilde{q}^{+}_{A}=\tilde{u}^{+I'}\tilde{q}_{AI'}$. This is a consistent truncation if $\{D_{A-},D_{B-}\}=\{\tilde{D}^A_{-},\tilde{D}^{B}_{-}\}=\{D_{A-},\tilde{D}^{B}_{-}\}=0$ so that one can consistently impose $D_{A-}\phi=\tilde{D}^{A}_{-}\phi=0$. This is true since 
$$\{D_{AI},D_{BJ}\}=C_{IJ}\partial_{AB},\,\,\{\tilde{D}^A_{I'},\tilde{D}^{B}_{J'}\}=C_{I'J'}\partial^{AB},$$
where $C_{IJ}$ is antisymmetric. Thus we will construct the on-shell superamplitude as a function of only $q^{A+},\tilde{q}^{+}_{A}$ or equivalently $\eta^+_a,\;\tilde{\eta}^+_{\dot{a}}$
$$M=M(p,\eta_a^+,\tilde{\eta}_{\dot{a}}^+).$$
From now on we drop the $+$ for simplicity.\footnote{R-symmetry is not really manifest since we do not integrate over the harmonics.} 
\begin{figure}
\begin{center}
\includegraphics[scale=0.8]{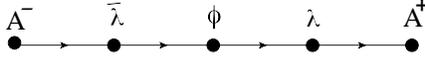}
\caption{The weight space diagram for 4D N=4 super Yang-Mills}
\end{center}
\end{figure}

The group theoretical interpretation of the $\eta$s is that they are the raising and lowering generators defined on the weight space of the little group \cite{Boels:2009bv}. For example, in 4-dimensions the physical states can be conveniently written as  states in the weight space of the U(1) little group fig.(1). A self-CPT spectrum then means that one has enough susy, and therefore enough $\eta$s, to reach all the physical states. Note that the lowering generators, represented by $\bar{\eta}$s, are absent. The fact that we began with $A^-$ reflects the fact that the on-shell superspace is a chiral superspace. In 6-dimensions the states now lie in the weight space of SU(2)$\times$SU(2), fig.(2).  
\begin{figure}
\begin{center}
\includegraphics[scale=0.9]{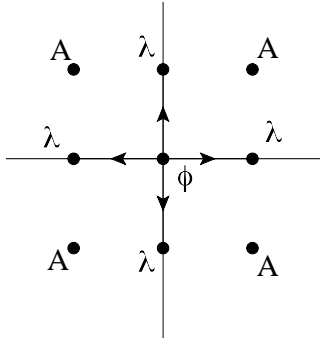}
\caption{The weight space diagram for 6D N=2 super Yang-Mills. Note that the gauginos are complex and there are two independent complex scalar field.}
\end{center}
\end{figure}
Using the 4 $\eta_{a},\tilde{\eta}_{\dot{a}}$s, one can begin with the scalar and reach all the other physical states.

For future reference we define the following fermionic delta functions
\eqa
\nonumber\delta^8(\sum _i q_i)=&&\left[\frac{1}{4!}\e_{ABCD}\delta(\sum_iq_i^A)\delta(\sum_jq_j^B)\delta(\sum_kq_k^C)\delta(\sum_lq_l^D)\right]\\
\nonumber\times&&\left[\frac{1}{4!}\e^{EFGH}\delta(\sum_i\tilde{q}_{iE})\delta(\sum_j\tilde{q}_{jF})\delta(\sum_k\tilde{q}_{kG})\delta(\sum_l\tilde{q}_{lH})\right]\\
\nonumber=&&\delta^4(\sum_i q_i^M)\delta^4(\sum_j \tilde{q}_{jM}),
\eqae
where the sum is over external legs. Notice the YM field strength appears as 
$$\int d(\eta_1)_{a}\int d(\tilde{\eta}_1)_{\dot{a}}\delta(\sum_iq_i^A)\delta(\sum_j\tilde{q}_{jB})=(\lambda_1)^A_a(\tilde{\lambda}_1)_{B\dot{a}}=(F^A\,_B)_{a\dot{a}}.$$
In this form it is then straight forward to supersymmetrize  Cheung and O'Connell's result.

Before going on to the super on-shell amplitudes, we would like to comment on the relationship to 4 dimensions off-shell superspace. For our purpose the precise nature of the harmonics $u$ and $\tilde{u}$ which paremeterize the double coset $\frac{SU(2)}{U(1)}\times \frac{SU(2)}{U(1)}$, is irrelevant for on-shell amplitudes. However, this R-coset space appears to be very similar to the projective superspace recently proposed for N=4 super Yang-Mills\cite{Hatsuda:2008pm}, this superspace is based on the supercoset $\frac{OSp(4|4)}{OSp(2|2)^2}$. If one uses the covering group, then the R-space part becomes 
$$\frac{SO(4)}{SO(2)^2}\rightarrow \frac{SU(2)}{U(1)}\times \frac{SU(2)}{U(1)}.$$
As we will see in the next section, the 4-point tree amplitude written in 6 dimensions has the same form as the 4-point amplitude derived in \cite{Hatsuda:2008pm}, in which the R-space parameters were evaluated at 0 anyway. The fact that the 6-dimensional on-shell amplitude shares the same form as the 4 dimensional off-shell is not surprising since on-shell in 6 dimensions simply restricts to the 6 dimensional lightcone. Projecting out the scale (projective light-cone) one has a 4 dimensional space where the vectors are not constrained to be null. The fact that one can extrapolate the 4 dimensional amplitude from a higher dimension on-shell counterpart is of great convenience. Recent advances in the evaluation of the S-matrix, which are usually only valid on-shell\footnote{For example the use of BCFW relies on the fact that the complex deformation only produces simple poles. If one looks at the off-shell amplitude, the shift will in general produce double poles. This will lead to residues that do not factorize into two tree amplitudes as the usual BCFW.}, can then be used to analyse 4 dimensional off-shell amplitudes which may give implications to an off-shell action, which is still lacking. Another application would be to use these off-shell amplitudes as an alternative to the recently proposed IR regularization scheme for N=4 super Yang-Mills \cite{Alday:2009zm}.

\section{Amplitudes in superspace}
\subsection{4-point amplitude}
We begin with the 4-point amplitude since the supersymmetrization is relatively straightforward. The 4-point amplitude for 6 dimensional Yang-Mills is  
\eqa
\nonumber 6\,D:\, M_4=&&\frac{-i\langle 1_a2_b3_c4_d\rangle[1_{\dot{a}}2_{\dot{b}}3_{\dot{c}}4_{\dot{d}}]}{st}\\
\nonumber=&&\frac{-i\e_{ABCD}(\lambda_1)_{a}^A(\lambda_2)_{b}^B(\lambda_3)_{c}^C(\lambda_4)_{d}^D\e^{EFGH}(\tilde{\lambda}_1)_{E\dot{a}}(\tilde{\lambda}_2)_{F\dot{b}}(\tilde{\lambda}_3)_{G\dot{c}}(\tilde{\lambda}_4)_{H\dot{d}}}{st}.
\eqae
Rewriting this in terms of field strengths using (\ref{def}),
\eq
6\,D: M_4=\frac{-i\e_{ABCD}\e^{EFGH}(F_1^A\,_E)_{a\dot{a}}(F_2^B\,_F)_{b\dot{b}}(F_3^C\,_G)_{c\dot{c}}(F_4^D\,_H)_{d\dot{d}}}{st}.
\label{4point}
\eqe
It is instructive to compare to the 4 dimensional result,
\eq
4\,D: M_4=\frac{i\langle12\rangle^4}{\langle12\rangle\langle23\rangle\langle34\rangle\langle41\rangle}\rightarrow \frac{i(f_1)^{\alpha\beta}(f_2)_{\alpha\beta}(\tilde{f}_3)^{\dot{\alpha}\dot{\beta}}(\tilde{f}_4)_{\dot{\alpha}\dot{\beta}}}{st}.
\eqe 
Note that the difference with 6 dimensions is simply the way the field strengths contracts their Lorentz indices. Again this is because the field strengths in 4 dimensions are (anti)chiral.

From (\ref{4point}) one can deduce the supersymmetric form: 
\eq
6\,D\, susy :\mathcal{M}_{4}=-i\frac{\delta^4(\sum q)\delta^4(\sum \tilde{q})}{st},
\label{4pointsusy}
\eqe
where $i,j,\cdot\cdot$ of the integration measure can be any of the external legs. Note that the little group indices are carried by the integration measure; different choice of measure represents different helicity configuration. The Yang-Mills amplitude corresponds to choosing $\prod_{i=1}^4d\eta_{ia}d\tilde{\eta}_{i\dot{a}}$ as the integration measure. This is also the case in 4 dimensions, where the on-shell super amplitude is  
\eq
4\,D\, susy: \mathcal{M}_{4}=i\frac{\delta^4(\sum\lambda^\alpha\eta)\delta^4(\sum\lambda_\alpha\eta)}{\langle12\rangle\langle23\rangle\langle34\rangle\langle41\rangle}.
\eqe 
Note the integration measure transforms under the U(1) little group.

One can compare (\ref{4pointsusy}) to the 4-point N=4 amplitude derived in \cite{Hatsuda:2008pm} 
\eq
4\,D\, projective: \mathcal{M}_{4}=i\int d\pi_i^{32}\frac{\delta^4(\sum\bar{\pi}^{a\dot{\alpha}})\delta^4(\sum\pi^{a'\alpha})}{st}\phi(1)\phi(2)\phi(3)\phi(4),
\eqe 
where $\phi$ is the scalar field strength and $\pi$s are the conjugate supermomenta of the 8 fermionic coordinates of $\frac{OSp(4|4)}{OSp(2|2)^2}$. Note that the bosonic Yang-Mills field strength also appears in similar fashion:
\eqa
6\,D\, susy:\int d(\eta_1)_{a}\int d(\tilde{\eta}_1)_{\dot{a}}(q_1)^A(\tilde{q}_{1B})=(F^A\,_B)_{a\dot{a}}\leftrightarrow \left.\pi_{a'\alpha}\pi_{b'\beta}\phi\right|=\eta_{a'b'}f_{\alpha\beta}.
\eqae

\subsection{3-point amplitude}
The 3-point amplitude vanishes on-shell in real Minkowski space, however it is non trivial in complex momentum space. Since our aim is to use BCFW as a systematic way of generating higher point amplitudes, we will proceed to compute it with complex momenta. Amplitudes should be written in terms of Lorentz invariants, however for the 3-point amplitude one has the problem of vanishing Lorentz invariants due to kinematic constraints: $p_i^2=0,\sum_{i=1}^3p_i=0\rightarrow (p_i\cdot p_j)=0$. In 4 dimensions this is solved by using complex momenta or going to split signature with real momenta, then $\lambda$ and $\tilde{\lambda}$ are no longer related and one can set either $\langle ij\rangle$ or $[ij]$ to zero but not both. In 6 dimensions one has  
$$p_i\cdot p_j=0\rightarrow (\lambda_i)^{Aa}(\lambda_i)^B_a(\bar{\lambda}_j)_{A\dot{a}}(\bar{\lambda}_j)^{\dot{a}}_B=\det\langle i_a| j_{\dot{a}}]=0.$$ 
i.e.the 2$\times$2 matrix $\langle1_a|2_{\dot{a}}]$ has rank 1. Therefore Cheung and O'Connell solved this by introducing SU(2) spinor variables for these bi-spinor matrices $\langle i|j]_{a{\dot{a}}}=u_{ia}\tilde{u}_{j\dot{a}}$.\footnote{We give their definitions and properties in appendix (\ref{equations})} To define their inverse, due to their presence in the denominator for the polarization vectors (\ref{polar}), one introduces variables $w_{ja}$ defined by $u_{a}w_{b}-u_{b}w_{a}=C_{ab}$. This definition defines $w_{ja}$ up to a shift $w_{ja}\rightarrow w_{ja}+b_ju_{ja}$. This ambiguity can be partially removed by requiring 
$$w_1^a\lambda^A_{1a}+w_2^a\lambda^A_{2a}+w_3^a\lambda^A_{3a}=0.$$
Then $w_i^a$ are defined up to shifts with $b_1+b_2+b_3=0$. Even though there is still ambiguity, this will help us determine the full amplitude by requiring invariance under this shift.

The 3-pt Yang-Mills amplitude is given as  
\eq
\nonumber6\,D\;M_{3}=i\Gamma_{abc}\tilde{\Gamma}_{\dot{a}\dot{b}\dot{c}}=i(u_1u_2w_3+u_1w_2u_3+w_1u_2u_3)_{abc}(\tilde{u}_1\tilde{u}_2\tilde{w}_3+\tilde{u}_1\tilde{w}_2\tilde{u}_3+\tilde{w}_1\tilde{u}_2\tilde{u}_3)_{\dot{a}\dot{b}\dot{c}}.\\
\eqe

To motivate the structure of the corresponding super amplitude, we cast the 3 point amplitude into the BCFW construction. Through BCFW, the 4-point amplitude can be constructed by sewing two 3-point amplitudes and integrating away 4 $\eta$s that carry the helicities of the propagator. Since the resulting 4-point amplitude has 8 fermionic delta functions, this requires the 3-point amplitudes to carry a total of 12 delta functions. Indeed in 4 dimensions, one is required to sew an MHV and an $\overline{{\rm MHV}}$ amplitude. Since $\overline{{\rm MHV}}$ has 8 delta functions in the anti-chiral $\bar{\eta}$, one Fourier transforms it into $\eta$s and results in a form that has only 4 delta functions, a total of 12. As discussed previously, in 6 dimensions there is no difference between MHV and $\overline{{\rm MHV}}$, while the number of $\eta$s to integrate remains the same. This leads to the conclusion that the 6 dimensions 3-point amplitude should be given with 6 delta functions and one has the following result:

\eqa
6\,D\,susy:\;\mathcal{M}_{3}=\frac{i}{2}\left[\delta(\sum q^A)\delta(\sum \tilde{q}_{A})\right]^2\delta(\sum w^{b}\eta_{b})\delta(\sum \tilde{w}^{\dot{b}}\tilde{\eta}_{\dot{b}}).
\eqae
To confirm this is true choose a specific piece of the integration measure, integrating $\eta^a_1\eta^b_2\eta^c_3\tilde{\eta}^{\dot{a}}_1$ $\tilde{\eta}^{\dot{b}}_2\tilde{\eta}^{\dot{c}}_3$. The combination of the form $[\eta^a_1\eta^b_2\tilde{\eta}^{\dot{a}}_1\tilde{\eta}^{\dot{b}}_2]\tilde{\eta}^{\dot{c}}_3\eta^c_3$ gives\footnote{ The brackets denote which of the $\eta$s are coming from the $\delta(q)$s}  
\eqa
\nonumber i\langle1_{a} |2_{\dot{b}}]\langle2_b |1_{\dot{a}}]w_{3c}\tilde{w}_{3\dot{c}}=-i\tilde{u}_{1\dot{a}}\tilde{u}_{2\dot{b}}u_{1a}u_{2b}w_{3c}\tilde{w}_{3\dot{c}},
\eqae
which would be one term in the YM expansion. Similarly if one integrates $\eta^a_2\eta^b_2\eta^c_1\tilde{\eta}^{\dot{a}}_1\tilde{\eta}^{\dot{b}}_2\tilde{\eta}^{\dot{c}}_3$ this gives
\eqa
\nonumber iu_{1c}(\tilde{u}_{1\dot{a}}\tilde{u}_{2\dot{b}}\tilde{w}_{3\dot{c}}+\tilde{u}_{1\dot{a}}\tilde{w}_{2\dot{b}}\tilde{u}_{3\dot{c}}+\tilde{w}_{1\dot{a}}\tilde{u}_{2\dot{b}}\tilde{u}_{3\dot{c}}).
\eqae
This is the amplitude for two gauginos and one gauge boson ($g_1,\tilde{\lambda}_2,\tilde{\lambda}_3$). Again this amplitude is invariant under the $b$ shift.

\subsection{5-point amplitude} 
The 5-point amplitude written in terms of field strengths and momenta is:
\eqa
\nonumber6\,D:\; M_5&=&\;\frac{i}{s_{12}s_{23}s_{34}s_{45}s_{51}}\left\{F_{1}^A\,_B(\displaystyle{\not}p_2\displaystyle{\not} p_3\displaystyle{\not} p_4\displaystyle{\not} p_5)_A\,^B\;(F_{2} \cdot F_{3}\cdot F_{4}\cdot F_{5})\right.\\
\nonumber&& +\frac{3}{10}\left[(\displaystyle{\not}p_2\displaystyle{\not} p_3\displaystyle{\not} p_4\displaystyle{\not} p_5)-(\displaystyle{\not}p_2\displaystyle{\not} p_5\displaystyle{\not}p_4\displaystyle{\not}p_3)\right]_A\,^B[F_1^A\,_DF_2^C\,_B(F_3 \cdot F_4 \cdot F_5)^D\,_C]\\
\nonumber&& +\frac{3}{10}\left[(\displaystyle{\not}p_2\displaystyle{\not} p_3\displaystyle{\not} p_4\displaystyle{\not} p_5)-(\displaystyle{\not}p_2\displaystyle{\not} p_5\displaystyle{\not} p_4\displaystyle{\not} p_3)\right]^A\,_B[F_1^C\,_AF_2^B\,_D(F_3 \cdot F_4 \cdot F_5)^D\,_C]\\
\nonumber &&+\frac{1}{10}\left[(\displaystyle{\not}p_5\displaystyle{\not} p_1\displaystyle{\not} p_2\displaystyle{\not} p_3)-(\displaystyle{\not}p_5\displaystyle{\not} p_3\displaystyle{\not} p_2\displaystyle{\not} p_1)\right]_A\,^B[F_3^A\,_DF_5^C\,_B(F_1 \cdot F_2 \cdot F_4)^D\,_C]\\
\nonumber &&+\frac{1}{10}\left.\left[(\displaystyle{\not}p_5\displaystyle{\not} p_1\displaystyle{\not} p_2\displaystyle{\not} p_3)-(\displaystyle{\not}p_5\displaystyle{\not} p_3\displaystyle{\not} p_2\displaystyle{\not} p_1)\right]^A\,_B[F_3^C\,_AF_5^B\,_D(F_1 \cdot F_2 \cdot F_4)^D\,_C]+cyclic\right\},\\
\eqae  
where $(\displaystyle{\not}p_2\displaystyle{\not} p_3\displaystyle{\not} p_4\displaystyle{\not} p_5)_A\,^E\equiv p_{2AB}p_3^{BC} p_{4CD}p_5^{DE}$, and we have dropped the SU(2) indices. Super symmetrizing we have:  
\eqa
\nonumber&&6\,D\;susy:\; \mathcal{M}_5=\;\frac{i\delta^4(\sum q)\delta^4(\sum\tilde{q})}{s_{12}s_{23}s_{34}s_{45}s_{51}}\big\{q_{1}(\displaystyle{\not}p_2\displaystyle{\not} p_3\displaystyle{\not} p_4\displaystyle{\not} p_5)\tilde{q}_{1}\;\big.\\
\nonumber&& +\frac{3}{10}q_{1}\left[(\displaystyle{\not}p_2\displaystyle{\not} p_3\displaystyle{\not} p_4\displaystyle{\not} p_5)-(\displaystyle{\not}p_2\displaystyle{\not} p_5\displaystyle{\not}p_4\displaystyle{\not} p_3)\right]\tilde{q}_{2} +\frac{3}{10}\tilde{q}_{1}\left[(\displaystyle{\not}p_2\displaystyle{\not} p_3\displaystyle{\not} p_4\displaystyle{\not} p_5)-(\displaystyle{\not}p_2\displaystyle{\not} p_5\displaystyle{\not} p_4\displaystyle{\not} p_3)\right]q_{2}\\
\nonumber &&+\frac{1}{10}q_{3}\left[(\displaystyle{\not}p_5\displaystyle{\not} p_1\displaystyle{\not} p_2\displaystyle{\not} p_3)-(\displaystyle{\not}p_5\displaystyle{\not} p_3\displaystyle{\not} p_2\displaystyle{\not} p_1)\right]\tilde{q}_{5}+\frac{1}{10}\left.\tilde{q}_{3}\left[(\displaystyle{\not}p_5\displaystyle{\not} p_1\displaystyle{\not} p_2\displaystyle{\not} p_3)-(\displaystyle{\not}p_5\displaystyle{\not} p_3\displaystyle{\not} p_2\displaystyle{\not} p_1)\right]q_5+cyclic\right\},\\
\label{5point}
\eqae  
where $q_{1}(\displaystyle{\not}p_2\displaystyle{\not}p_3\displaystyle{\not}p_4\displaystyle{\not}p_5)\tilde{q}_{1}=q^M_{1}(\displaystyle{\not}p_2\displaystyle{\not}p_3\displaystyle{\not}p_4\displaystyle{\not}p_5)_M\,^N\tilde{q}_{1N}$.

\section{BCFW construction}
Here we give a short introduction to the BCFW construction and show how to obtain our 4-point result from the 3-point. We begin by shifting the momenta of two arbitrary external lines, say 1 and 2, by a vector $q$:
$$\hat{p}_1=p_1+zq,\;\hat{p}_2=p_2-zq.$$
We require the vector $q$ to satisfy 
$$q^2=q\cdot p_1=q\cdot p_2=0,$$
so that the deformed momenta remain on-shell, $\hat{p}^2_1=\hat{p}^2_2=0$. This can be done by choosing $q$ to be related to the polarization of line 1, $q\sim\e_1$, and choosing $\lambda_2$ as the reference spinor $\mu$. However the polarization vector has additional little group index. One remedies this by contracting it with an auxiliary parameter $x^{a\dot{a}}$\cite{Cheung:2009dc}
$$q^{AB}=x^{a\dot{a}}\frac{\left|^{[A}1_{a}\rangle\langle2_{c}\,^{B]}\right|}{\langle2_{c}|1^{\dot{a}}]}.$$ 
Then the requirement of $q^2=0$ implies $\det x^{a\dot{a}}=0$, i.e, $x^{a\dot{a}}=x^{a}\tilde{x}^{\dot{a}}$. Since the amplitude is a rational function of momentum, this deformation will result in a complex function with only simple poles. The poles are the propagators in the denominator and are simple since 
$$\hat{P}_{1j}^2=(\hat{p}_1+\cdot\cdot\cdot p_j)^2=P^2_{1j}+z2q\cdot P_{1j}\rightarrow z_{1j}=-\frac{P^2_{1j}}{2q\cdot P_{1j}},$$
where $P_{1j}$ represents the sum of momentum on one side of the propagator. Note that if the shifted lines are either not included or both included, one will not develop a pole and the corresponding graphs do not contribute. 

If the complex amplitude vanishes as $z\rightarrow \infty$, then it is uniquely determined by it's residues:
$$A(z)=\sum_{j}\frac{c_{1j}}{z-z_{1j}}.$$
Our physical amplitude then corresponds to $A(0)=-\sum_{j}\frac{c_{1j}}{z_{1j}}$. The sum is understood as summing different ways of separating the amplitude in two halves with the propagator producing the pole. The residues $c_{1j}$ take the form
\eq
c_{1j}=-\hat{A}_L\times \hat{A}_R\frac{1}{2q\cdot P_{1j}}\left|_{z=\frac{P^2_{1j}}{2q\cdot P_{1j}}}\right. ,
\eqe
and therefore 
$$A(0)=\sum_{j}\hat{A}_L(\hat{p}_1,\cdot\cdot\cdot,p_j,-\hat{P}_{1j})\frac{i}{P_{1j}^2}\hat{A}_R(\hat{P}_{1j},\cdot\cdot, \hat{p}_2)\left|_{z=\frac{P^2_{1j}}{2q\cdot P_{1j}}}\right. .$$
Since both $\hat{p}_1,\hat{p}_2$ are on-shell and $\hat{P}_{1j}$ is also on-shell when the shift is evaluated at the pole, each function on either side of the propagator becomes itself an on-shell amplitude of lower points. Thus BCFW expresses an $n$ point amplitude in terms of lower point on-shell amplitudes with two of their external momenta deformed. 

An important ingredient is the fact that $A(z)$ vanishes as $z\rightarrow\infty$, this is true for maximal supersymmetric theories in 4 dimensions and general pure gauge and gravity theories \cite{ArkaniHamed:2008gz,ArkaniHamed:2008yf}. Since the 3 and 4-point amplitudes have only delta functions in the numerator, if one shifts in a way that preserves the (super)momentum conservation relation, the amplitudes automatically vanishes at large $z$. Indeed we define our super symmetric shifts to satisfy these conditions as we show in appendix (\ref{susyshift}). For higher point amplitudes the numerator will have, besides the (super)momentum conservation delta functions, individual $\displaystyle{\not}p_i$s and $q_i$s. The reason why one might produce more $z$s in the numerator than the purely Yang-Mills case, is the integration of $\eta_1^2$ or $\eta_2^2$ which correspond to non-vector reference lines. From the form of  the shifted $q_{\hat{1}},q_{\hat{2}}$ in (\ref{combo}), one can see that integrating $\eta_1^2$ or $\eta_2^2$ will not produce $z^2$ terms. Therefore non-vector reference lines will only produce shifts in $q_i$s that are at most linear in $z$, which is the same degree as purely Yang-Mills. Thus we conclude that in principle the supersymmetric theory should vanish at large $z$ if the Yang-Mills theory vanishes. Note that our argument is similar to the on-shell supersymmetric Ward identity used in \cite{ArkaniHamed:2008gz}.
\subsection{BCFW for 4-point}
\begin{figure}
\begin{center}
\includegraphics[scale=0.8]{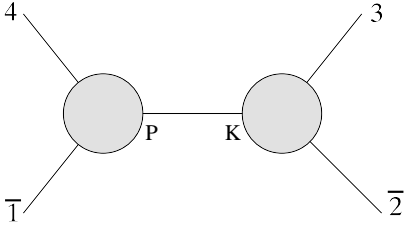}
\end{center}
\caption{The 4-point amplitude in the BCFW formalism. This is the only graph contributing if one chooses $1$ and $2$ as the shifted legs.}
\end{figure}
Now let us compute the 4-point amplitude. Choosing 1 and 2 as the shifted leg, the only graph that will be contributing will be the $t$ channel graph fig.(3) 

Then the BCFW for super Yang-Mills is written as
\eqa
\nonumber &&\frac{-1}{4}\left[\frac{1}{2}\int d\eta^a_P\int d\tilde{\eta}_{P}^{\dot{a}}\right]\left[\frac{1}{2}\int d\eta_{Pa}\int d\tilde{\eta}_{P\dot{a}}\right]\left[\delta(\sum_L q^A)\delta(\sum_L\tilde{q}_{A})\right]^2[\delta(\sum_L w_n^{b}\eta_{nb})\delta(\sum_L \tilde{w}_{n\dot{b}}\tilde{\eta}_{n}^{\dot{b}})]\\
\nonumber&&\times\frac{i}{t}\left[\delta(\sum_R q^B)\delta(\sum_R \tilde{q}_{B})\right]^2[\delta(\sum_R w_n^{c}\eta_{nc})\delta(\sum_R \tilde{w}_{n\dot{c}}\tilde{\eta}_{n}^{\dot{c}})],
\eqae 
where the $\eta_p$ integrals essentially keep track of the helicity in the propagator. The fermionic delta functions are explicitly 
\eqa
\nonumber&&\delta(\sum_L q^{A})=\delta( \lambda_{\hat{1}}^{Aa}\eta_{\hat{1}a}+\lambda_4^{Aa}\eta_{4a}+\lambda_P^{Aa}\eta_{Pa}),\;\delta(\sum_R q^{A})=\delta( \lambda_{\hat{2}}^{Aa}\eta_{\hat{2}a}+\lambda_3^{Aa}\eta_{3a}-\lambda_P^{Aa}\eta_{Pa})\\
\nonumber&&\delta(\sum_L w_i^{b}\eta_{ib})=\delta(w_{\hat{1}}^{b}\eta_{\hat{1}b}+w_4^{b}\eta_{4b}+w_P^{b}\eta_{Pb}),\;\delta(\sum_R w_i^{b}\eta_{ib})=\delta(w_{\hat{2}}^{b}\eta_{\hat{2}b}+w_3^{b}\eta_{3b}+iw_K^{b}\eta_{Pb}).\\
\eqae
The spinors $\lambda_K$ ($\lambda_P$) is defined from $p_K=-\hat{p}_2-p_3$ ($p_P=-\hat{p}_1-p_4$) which are on-shell due to the shift. One then integrates over the $\eta_P$s. There are three different ways of picking up two $\eta_{P}$s from    
\eqa
\nonumber&&\delta(\sum_R w_i^{b}\eta_{ib})\delta(\sum_L w_j^{c}\eta_{jc})\delta^2( \lambda_{\hat{1}}^{Aa}\eta_{\hat{1}a}+\lambda_4^{Aa}\eta_{4a}+\lambda_P^{Aa}\eta_{Pa})\delta^2( \lambda_{\hat{2}}^{Bd}\eta_{\hat{2}d}+\lambda_3^{Bd}\eta_{3d}-\lambda_P^{Bd}\eta_{Pd})\\
\nonumber=&&\delta^2( \sum_{full} q^A)\delta(\sum_R w_i^{a}\eta_{ia})\delta(\sum_L w_j^{b}\eta_{jb})\delta^2( \lambda_{\hat{2}}^{Bc}\eta_{\hat{2}c}+\lambda_3^{Bc}\eta_{3c}-\lambda_P^{Bc}\eta_{Pc}).
\eqae
One can either choose both $\eta_P$ from $\delta w$, one from $\delta w$ and one from $\delta q^A$ and finally taking both from $\delta q^A$. The last two way give vanishing results since they produce terms proportional to a $\lambda_P^A$. These terms contract with either $p_{\hat{2}}+p_{3}$ or $(\tilde{\lambda}_{\hat{2}}\cdot\tilde{\eta}_{\hat{2}}+\tilde{\lambda}_{3}\cdot\tilde{\eta}_{3})_A$, which vanish either due to momentum conservation or the fermionic delta function. Therefore integrating over $\eta_P$ gives 
\eqa
\nonumber&&\frac{i}{t}\delta^4(\sum_{i=1}^4 q^A)\delta^4(\sum_{j=1}^4 \tilde{q}_{A})w_{P}^{d}\tilde{w}_{K\dot{d}}w_{Kd} \tilde{w}_{P}^{\dot{d}}\\
\nonumber=&&\frac{-i}{st}\delta^4(\sum_{i=1}^4 q^A)\delta^4(\sum_{j=1}^4 \tilde{q}_{A})
\eqae
where in the last line we've used $w_{P}^{d}\tilde{w}_{K\dot{d}}w_{Kd} \tilde{w}_{P}^{\dot{d}}=-\frac{1}{s}$, we will demonstrate this in appendix (\ref{equations}). 

\subsection{BCFW for 5-point}
There are two contributions to the 5-point amplitude as shown in fig.(4). Now the crucial point is that the auxiliary parameter $x,\tilde{x}$ introduced by the shift should cancel out in the end. This should be automatic since the $x$s enter the BCFW with $z$s, good large $z$ behaviour then automatically ensures they drop out in the end. Explicitly showing this will produce a final result that is in a compact form. In the Yang-Mills computation, these parameters cancel after combining the two graphs $D_1,D_2$. In principle the BCFW for super amplitude should be parallel to the Yang-Mills calculation, since the only difference is the integrating of the $\eta_P$s that carry the degrees of freedom in the propagator. Here we will follow suit and compute the two graphs separately, after performing the integration we will show that the result has the same form as Yang-Mills and therefore the cancellation goes through accordingly and one can read off the supersymmetric result straight forwardly.
\begin{figure}
\begin{center}
\includegraphics[scale=0.9]{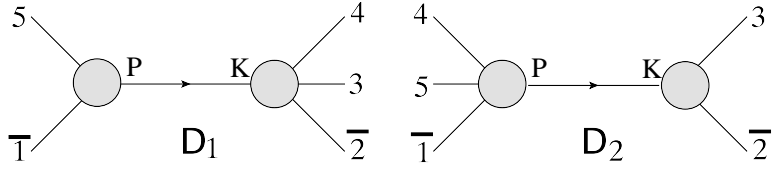}
\end{center}
\caption{The 5-point amplitude in the BCFW formalism. Now there are two graphs contributing if one chooses $1$ and $2$ as the shifted legs.}
\end{figure}

\noindent ${\bf D_1}$:

We compute
\eqa
\nonumber D_1=\frac{i}{2}\frac{1}{s_{51}\hat{s}_{23}s_{34}}\int d^4\eta_P \delta^4(\sum_R q^A)\delta^4(\sum_R \tilde{q}_A) \left[\delta(\sum_L q^B)\delta(\sum_L \tilde{q}_B)\right]^2\delta(\sum_L w\eta)\delta(\sum_L \tilde{w}\tilde{\eta})\\
\nonumber=\frac{i}{2}\frac{1}{s_{51}\phi s_{34}}(q\cdot p_5)\delta^4(\sum_{full} q^A)\delta^4(\sum_{full} \tilde{q}_A)\int d^4\eta_P\left[ \delta(\sum_L q^B)\delta(\sum_L \tilde{q}_B)\right]^2\delta(\sum_L w\eta)\delta(\sum_L \tilde{w}\tilde{\eta}),
\eqae
where 
\eq
\hat{s}_{23}=2\hat{p}_2\cdot p_3=\frac{\phi}{q\cdot p_5},\;\;\phi=s_{23}q\cdot p_5+s_{51}q\cdot p_3.
\eqe
We've used that $z$ is evaluated at the pole $z=-\frac{s_{51}}{2q\cdot p_5}=-\frac{s_{51}s_{12}}{[\tilde{x}|p_5 p_2|x\rangle}$ for this graph. 

Now we do the $\eta_p$ integral. One observes that there must be at least one $\eta_p$ coming from the $w$ delta function, therefore the integrand becomes 
\eqa
\nonumber(w_k\cdot\lambda_k^A)(\tilde{w}_k\cdot\tilde{\lambda}_{kB})(\tilde{\lambda}_{\hat{1}A}\cdot\tilde{\eta}_{\hat{1}}+\tilde{\lambda}_{5A}\cdot\tilde{\eta}_5)(\lambda^B_{\hat{1}}\cdot\eta_{\hat{1}}+\lambda^B_{5}\cdot\eta_5)\\
=-(\tilde{u}_{\hat{1}}\cdot\tilde{\eta}_{\hat{1}}-\tilde{u}_{5}\cdot\tilde{\eta}_5)(u_{\hat{1}}\cdot\eta_{\hat{1}}-u_{5}\cdot\eta_5),
\label{5.7}
\eqae
where we used the fact that $[i|j\rangle$ on the three point vertex can be rewritten in terms of $u$ and $w$. Then we need to get rid of $u$s. Note that 
\eq
(q\cdot p_5)=\frac{x^{a} x^{\dot{a}}}{s_{12}}[5^{\dot{c}}|\hat{1}_a\rangle\langle2_b|5_{\dot{c}}]\langle2^b|\hat{1}_{\dot{a}}]=\frac{x^{a} x^{\dot{a}}}{s_{12}}\tilde{u}_5^{\dot{c}}u_{\hat{1}a}\langle2_b|5_{\dot{c}}]\langle2^b|\hat{1}_{\dot{a}}].
\label{5.8}
\eqe
Putting (\ref{5.7},\ref{5.8}) together, $D_1$ becomes\footnote{We give the derivation of (\ref{needproof}) and (\ref{lake}) in detail in appendix (\ref{proof})}
\eqa
\nonumber D_1&\sim&-i\frac{1}{s_{51}\phi s_{34}s^2_{12}}\left(\langle x|\displaystyle{\not}p_2\displaystyle{\not}p_5|\hat{1}_{\dot{a}}]\tilde{\eta}^{\dot{a}}_{\hat{1}}-s_{12}\langle x|5^{\dot{b}}]\tilde{\eta}_{5\dot{b}}\right)\left([\tilde{x}|\displaystyle{\not}p_2\displaystyle{\not}p_5|\hat{1}_c\rangle\eta^c_{\hat{1}}-s_{12}[\tilde{x}|5^d\rangle\eta_{5d}\right),\\
\label{needproof}
\eqae
where $\sim$ means dropping delta functions. Putting in the definition of the shifted quantities (\ref{combo}), one has 
\eqa
\nonumber& D_1&\sim- i\frac{1}{s_{51}\phi s_{34}s^2_{12}}\\
\nonumber&\times&\left[\langle x|\displaystyle{\not}p_2\displaystyle{\not}p_5|1]\cdot\tilde{\eta}_{1}-z\langle x|\displaystyle{\not}p_2\displaystyle{\not}p_5|\tilde{x}][2^{\dot{b}}|x\rangle\tilde{\eta}_{2\dot{b}}/s_{12}+z\langle x|2^{\dot{c}}]\langle x|\displaystyle{\not}p_2\displaystyle{\not}p_5|2_{\dot{c}}]\tilde{x}^{\dot{a}}\tilde{\eta}_{1\dot{a}}/s_{12}-s_{12}\langle x|5^{\dot{d}}]\tilde{\eta}_{5\dot{d}}\right]\\
\nonumber&\times&\left[[\tilde{x}|\displaystyle{\not}p_2\displaystyle{\not}p_5|1\rangle\cdot\eta_{1}-z[\tilde{x}|\displaystyle{\not}p_2\displaystyle{\not}p_5|x\rangle[\tilde{x}|2^b\rangle\eta_{2b}/s_{12}+z[\tilde{x}|2^b\rangle[\tilde{x}|\displaystyle{\not}p_2\displaystyle{\not}p_5|2_b\rangle x^a\eta_{1a}/s_{12}-s_{12}[\tilde{x}|5^d\rangle\eta_{5d}\right].
\eqae
Using  $[\tilde{x}|\displaystyle{\not}p_2|\tilde{x}]=\tilde{x}^{\dot{a}}\tilde{x}^{\dot{b}}[1_{[\dot{a}}|\displaystyle{\not}p_2|1_{\dot{b}]}]=0$ and substituting the value of $z$, one then arrives at 
\eqa
\nonumber D_1&=&-i\frac{\delta^4(\sum_{full} q^A)\delta^4(\sum_{full} \tilde{q}_B)}{s_{51}\phi s_{34}s^2_{12}}\left[\langle x|\displaystyle{\not}p_2\displaystyle{\not}p_5|1]\cdot\tilde{\eta}_{1}+[2^{\dot{b}}|x\rangle\tilde{\eta}_{2\dot{b}}s_{51}-s_{12}\langle x|5^{\dot{d}}]\tilde{\eta}_{5\dot{d}}\right]\\
\nonumber&&\times\left[[\tilde{x}|\displaystyle{\not}p_2\displaystyle{\not}p_5|1\rangle\cdot\eta_{1}+[\tilde{x}|2^b\rangle\eta_{2b}s_{51}-s_{12}[\tilde{x}|5_d\rangle\eta_5^d\right].\\
\label{D1}
\eqae

\noindent ${\bf D_2}$:

For the second graph, we compute:
\eqa
\nonumber D_2=\frac{i}{2}\frac{(q\cdot p_3)}{s_{23}\phi s_{45}} \delta^4(\sum_{full} q^A)\delta^4(\sum_{full} \tilde{q}_B)\int d^4\eta_P\left[ \delta(\sum_L q^B)\delta(\sum_L \tilde{q}_B)\right]^2\delta(\sum_L w\eta)\delta(\sum_L \tilde{w}\tilde{\eta}).
\eqae

After integrating $\eta_p$, $D_2$ is proportional to $(q\cdot p_3)(-\tilde{u}_{\hat{2}}\cdot\tilde{\eta}_{\hat{2}}+\tilde{u}_{3}\cdot\tilde{\eta}_3)(u_{\hat{2}}\cdot\eta_{\hat{2}}-u_{3}\cdot\eta_3)$, which again we would need to combine in order to covert the $u,\tilde{u}$. We begin with  
\eqa
\nonumber(q\cdot p_3)&=&-\frac{x^{a} x^{\dot{a}}}{s_{12}}\langle1_a|\displaystyle{\not}p_3\displaystyle{\not}p_2|1_{\dot{a}}]=-\frac{x^{a} x^{\dot{a}}}{s_{12}}\langle1_a|\displaystyle{\not}p_3\displaystyle{\not}\hat{p}_2|1_{\dot{a}}]\\
\nonumber&=&-\frac{x^{a} x^{\dot{a}}}{s_{12}}(\tilde{u}_3\cdot[3|1_a\rangle)( u_{\hat{2}}\cdot\langle\hat{2}|1_{\dot{a}}])=\frac{-x^{a} x^{\dot{a}}}{s_{12}}\tilde{u}_{\hat{2}\dot{c}}[\hat{2}^{\dot{c}}|1_a\rangle u_{\hat{2}b}\langle\hat{2}^b|1_{\dot{a}}],
\eqae
where we've used $\tilde{u}_3\cdot[3|=\tilde{u}_{\hat{2}}\cdot[\hat{2}|$. Again using $u_i\tilde{u}_j=\langle i|j]$, $D_2$ becomes
\eqa
\nonumber D_2&\sim&-i\frac{1}{s_{23}\phi s_{45}s_{12}}\left[[\tilde{x}|\displaystyle{\not}p_3|\hat{2}_{\dot{a}}]\tilde{\eta}^{\dot{a}}_{\hat{2}}+[\tilde{x}|\displaystyle{\not}p_2|3_{\dot{a}}]\tilde{\eta}^{\dot{a}}_{3}\right]\left[\langle x|\displaystyle{\not}p_3|\hat{2}_b\rangle\eta_{\hat{2}}^b+\langle x|\displaystyle{\not}p_2|3_b\rangle\eta_{3}^b\right].\\
\label{lake}
\eqae
Again using the form of the shifted quantities (\ref{combo}): 
\eqa
\nonumber&&-i\frac{1}{s_{23}\phi s_{45}s_{12}}\left[-[\tilde{x}|\displaystyle{\not}p_3|2^{\dot{d}}]\tilde{\eta}_{2\dot{d}}-z[\tilde{x}|\displaystyle{\not}p_3\displaystyle{\not}p_2|x\rangle\tilde{x}^{\dot{b}}\tilde{\eta}_{1\dot{b}}/s_{12}-z[\tilde{x}|\displaystyle{\not}p_3|\tilde{x}]\langle x|2^{\dot{a}}]\tilde{\eta}_{2\dot{a}}/s_{12}+[\tilde{x}|\displaystyle{\not}p_2|3_{\dot{d}}]\tilde{\eta}^{\dot{d}}_{3}\right]\\
\nonumber&&\times\left[-\langle x|\displaystyle{\not}p_3|2^d\rangle\eta_{\hat{2}d}-z\langle x|\displaystyle{\not}p_3\displaystyle{\not}p_2|\tilde{x}] x^b\eta_{1b}/s_{12}-z\langle x|\displaystyle{\not}p_3|x\rangle[\tilde{x}|2^a\rangle\eta_{2a}/s_{12}+\langle x|\displaystyle{\not}p_2|3_b\rangle\eta_{3}^b\right]\\
\nonumber&=&-i\frac{1}{s_{23}\phi s_{45}s_{12}}\left([\tilde{x}|\displaystyle{\not}p_3|2^{\dot{d}}]\tilde{\eta}_{2\dot{d}}+s_{23}x^b\eta_{1b}+[\tilde{x}|\displaystyle{\not}p_2|3^{\dot{d}}]\tilde{\eta}_{3\dot{d}}\right)\left(\langle x|\displaystyle{\not}p_3|2^d\rangle\eta_{2d}+s_{23}x^b\eta_{1b}+\langle x|\displaystyle{\not}p_2|3^b\rangle\eta_{3b}\right),\\
\label{D2}
\eqae

Combining (\ref{D1}) and (\ref{D2}) we see that we've reproduced part of the result of Yang-Mills in \cite{Cheung:2009dc}, more precisely eq.(7.6) and (7.5). One can see the remaining part comes from the fermionic delta function $ \delta^4(\sum_{full} q^A)\delta^4(\sum_{full} \tilde{q}_B)$, if one chooses the purely Yang-Mills measure $d\eta_{1a}d\eta_{2b}d\eta_{3c}d\eta_{4d}d\eta_{5e}$ and $d\tilde{\eta}_{1\dot{a}}d\tilde{\eta}_{2\dot{b}}d\tilde{\eta}_{3\dot{c}}d\tilde{\eta}_{4\dot{d}}d\tilde{\eta}_{5\dot{e}}$. Therefore the remaining calculation resembles Yang-Mills case with the Schouten identity replaced by $q^E(\epsilon_{ABCD}q^Aq^Bq^Cq^D)=0$, i.e. the totally antisymmetric 5 index tensor vanishes if $A=1,\cdot\cdot,4$.  The result is (\ref{5point}).

\section{One-loop 4-point}
To show the power of this on-shell superspace, here we compute the one-loop 4-point amplitude for 6 dimensions maximal super Yang-Mills. It was shown in D dimensions maximal super Yang-Mills that the two-particle cut for the one-loop 4-point amplitude has the following relation\cite{Bern:1997nh}
$$\sum_{s_1,s_2}A_{{\rm tree}}(k_2^{s_2},1,2,-k_1^{s_1})A_{{\rm tree}}(-k_2^{s_2},3,4,k_1^{s_1})=-istA_{{\rm tree}}(1,2,3,4)\frac{1}{(p_1-k_1)^2(p_3-k_2)^2}$$
\begin{figure}
\begin{center}
\includegraphics[scale=0.9]{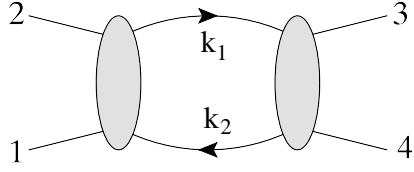}
\caption{Two-particle cut for one-loop 4-point amplitude.}
\end{center}
\end{figure}
where $s_1,s_2$ labes the internal states and are summed over. We now reproduce this relation in 6 dimensions. Using superspace to sum the internal states: 
\eqa
\nonumber&&\sum_{s_1,s_2}A_{{\rm tree}}(k_2^{s_2},1,2,-k_1^{s_1})A_{{\rm tree}}(-k_2^{s_2},3,4,k_1^{s_1})\\
\nonumber =&&-\int d^2\eta_{k_1}\int d^2\eta_{k_2}\int d^2\tilde{\eta}_{k_2}\int d^2\tilde{\eta}_{k_2}\frac{\delta^4(\sum_R q^A)\delta^4(\sum_R\tilde{q}_A)}{(p_1-k_1)^2s}\frac{\delta^4(\sum_L q^B)\delta^4(\sum_L\tilde{q}_B)}{s(p_3-k_2)^2}\\
\nonumber=&&-\int d^2\eta_{k_1}\int d^2\eta_{k_2}\int d^2\tilde{\eta}_{k_2}\int d^2\tilde{\eta}_{k_2}\frac{\delta^4(\sum_{full}q^A)\delta^4(\sum_{full}\tilde{q}_A)}{(p_1-k_1)^2s}\frac{\delta^4(\sum_L q^B)\delta^4(\sum_L\tilde{q}_B)}{s(p_3-k_2)^2}\\
\nonumber=&&-\frac{\delta^4(\sum_{full}q^A)\delta^4(\sum_{full}\tilde{q}_A)}{(p_1-k_1)^2s}\frac{(k_1\cdot k_2)^2}{s(p_3-k_2)^2}=-istA_{{\rm tree}}(1,2,3,4)\frac{1}{(p_1-k_1)^2(p_3-k_2)^2}
\eqae 
where we used $k_1-k_2=p_1+p_2$.

\section{Conclusion}
In this paper we present an on-shell superspace for maximal supersymmetric on-shell amplitudes in 6 dimensions. Combined with unitarity methods, one can efficiently study quantum corrections for 6D gauge and gravity theories. For example, this has potential application for studying the UV divergences of maximal supergravity at 4 loop where the critical dimension for finiteness is 5.5 \cite{Bern:2006kd}. This can also be used to study the N=4 theory near D=4 in the context of AdS/CFT. For non supersymmetric theories, one can also use these 6 dimensions tree amplitudes for constructing loop amplitudes using unitarity methods. The particles across the cuts are 6 dimensions and therefore may produce non-vanishing rational terms that were undetected using 4 dimensions tree amplitudes. One then sets the external lines to be in the 4 dimensions subspace in the end. The 6 dimensional spinors constructed here should also be useful in representing the S-matrix for the N=(2,0) theory. 

The other important feature is its close relation to 4 dimensions N=4 off-shell superspace. Being off-shell in 4 dimensions, this should provide a more suitable space to study the recently discovered dual superconformal symmetry\cite{Drummond:2008vq}, which is broken by IR singularities.     
\acknowledgments
We would like to thank Donal O'Connell and Clifford Cheung for detailed discussions of their six dimensional Yang-Mills calculation and John Joseph M. Carrasco for discussions on general structure of loop amplitudes. YH would especially like to thank Zvi Bern for the many suggestions and illuminating the potential applications in the preparation of this manuscript. The research of WS is supported by NSF grant No. PHY-0653342, and the research of YH is supported by DOE under contract DE-FG03-91ER40662. YH would also like to thank support from the National Center for Theoretical Sciences, Taiwan, R.O.C. during the completion of this work.  
\appendix
\section{6D spinors }
\label{4D}
\subsection{6 dimensional spinors in terms of momenta}
One of the interesting applications of the results presented here is to compute D dimensional cuts for the 4 dimensional theory. For this purpose, it is convenient to have a dictionary from which our 6 dimensional Lorentz invariants, written in terms of spinors variables, can be rewritten in terms of 6 dimensional momenta. 

Since a 6 dimensional vector is in the anti-symmetric representation of SU(4), the off-diagonal block of this 4$\times$4 matrix is then 4 dimensional. To make contact with the usual 4 dimensional notations we parameterize this off-diagonal 2$\times$2 block by $\sigma$ matrices
\eqa
\nonumber\Sigma^{\mu}_{(6)AB}=\left(\begin{array}{cc}0 & (\sigma^\mu)^{\alpha}\,_{\dot{\alpha}} \\-(\sigma^{\mu T})_{\dot{\alpha}}\,^\alpha & 0\end{array}\right)\,,\;\; \tilde{\Sigma}^{AB}_{(6)\mu}=\left(\begin{array}{cc}0 & (\sigma_\mu)_{\alpha}\,^{\dot{\alpha}} \\-(\sigma_{\mu}^{T})^{\dot{\alpha}}\,_\alpha & 0\end{array}\right),\;\;{\rm for}\;\mu=0,1,2,3\\
\eqae 
the $\sigma$ matrices are defined as usual: $\sigma^0=\left(\begin{array}{cc}-1 & 0 \\0 & -1\end{array}\right),\,\sigma^1=\left(\begin{array}{cc}0 & 1 \\1 & 0\end{array}\right),\,\sigma^2=\left(\begin{array}{cc}0 & -i \\i & 0\end{array}\right),\,\sigma^3=\left(\begin{array}{cc}1 & 0 \\0 & -1\end{array}\right)$. Note that the above matrices are equivalent to the 4 dimensional gamma matrices in the Weyl representation, i.e. $\Sigma^{\mu}_{(6)}=\gamma_{(4)}^\mu=\left(\begin{array}{cc}0 & (\sigma^\mu)^{\alpha}\,_{\dot{\alpha}} \\(\bar{\sigma}^\mu)_{\dot{\alpha}} \,^{\alpha}& 0\end{array}\right)$ for $\mu=0,1,2,3$. One also has 
$$\Sigma^{5}_{(6)AB}=\left(\begin{array}{cc}iC^{\alpha\beta} & 0 \\0 & iC^{\dot{\alpha}\dot{\beta}}\end{array}\right),\;\;\Sigma^{6}_{(6)AB}=\left(\begin{array}{cc}C^{\alpha\beta} & 0 \\0 & C^{\dot{\alpha}\dot{\beta}}\end{array}\right).$$

Now we explicitly solve the Dirac equation with generic 6 dimensional on-shell momenta;
\eq
\displaystyle{\not}k_{AB}\lambda^B_a=\left(\begin{array}{cc}\delta_{\alpha}\,^{\beta}(k_6+ik_5) & k_\mu\sigma^\mu\,_{\alpha\dot{\alpha}} \\k_\mu\bar{\sigma}^{\mu\dot{\beta}\beta} & (k_6-ik_5)\delta^{\dot{\beta}}\,_{\dot{\alpha}}\end{array}\right)\left(\begin{array}{c}\lambda_{\beta} \\ \lambda^{\dot{\alpha}}\end{array}\right)_a=0,\quad\mu=0,1,2,3.
\eqe
We have split the 6 dimensions SU(4) spinor in half, $\lambda^A=(\lambda_\alpha,\;\lambda^{\dot{\alpha}})$, since it is desirable to stay as close to the well known 4 dimensional spinor as possible. The solution has been constructed by Boels \cite{Boels:2009bv}, here we summarize the results. One start by writting ($k_{(4)}^\mu=k^0,k^1,k^2,k^3$) in terms of two spinors
\eq
k_{(4)\alpha\dot{\alpha}}=k_\alpha k_{\dot{\alpha}}+\frac{k_{(4)}^2}{2q\cdot k}q_\alpha q_{\dot{\alpha}}
\eqe 
where $q_{\alpha\dot{\alpha}}$ is again an arbitrary null vector with $q\cdot k\neq0$. One sees that $k_\alpha,\; k_{\dot{\alpha}}$, are the 4 dimensional spinors associated to the shifted 4 dimensional momenta. Then the solution to the Dirac equation is a $4\times2$ matrix reflecting the two dimensional space of solution.     
\eq
\lambda^{A}_a=\left(\begin{array}{cc}(k_6-ik_5)\frac{q_\alpha}{\langle qk\rangle}& k_\alpha \\k^{\dot{\alpha}}& (k_6+ik_5)\frac{q^{\dot{\alpha}}}{[qk]}\end{array}\right)
\eqe
and similarly 
\eq
\tilde{\lambda}_{A\dot{a}}=\left(\begin{array}{cc}(k_6+ik_5)\frac{q^\alpha}{\langle qk\rangle}& k^\alpha \\k_{\dot{\alpha}}& (k_6-ik_5)\frac{q_{\dot{\alpha}}}{[qk]}\end{array}\right)
\eqe
Again since the 4 dimensional spinor inner products can be expressed in terms of momenta, all our Lorentz invariants can be expressed in terms of momenta. 
If one constructs higher-point amplitude through BCFW construction, sometimes it might be preferable not to factorize out all the shifted variables. This would then leave behind SU(2) spinors $w^a,\,\tilde{w}^{\dot{a}}$. We properly define these SU(2) spinors in appendix (\ref{equations}), so their dependence on momentum can be easily derived from the above spinors.  
\subsection{6 dimensional spinors in terms of 4 dimensions}
Suppose all external momenta lie in a 4 dimensions subspace, one should then be able to extract the 4 dimensional amplitude from our 6 dimensional result. 
Setting $k_6=k_5=0$ the above solutions become
\eqa
\nonumber&&\lambda^A_{1}=\left(\begin{array}{c}0 \\ k^{\dot{\alpha}}\end{array}\right),\;\lambda^A_{2}=\left(\begin{array}{c}k_\alpha \\ 0\end{array}\right)\\
\nonumber &&\tilde{\lambda}_{A1}=\left(\begin{array}{c}0 \\ k_{\dot{\alpha}}\end{array}\right),\;\tilde{\lambda}_{A2}=\left(\begin{array}{c}k^\alpha \\ 0\end{array}\right).\\
\label{cusp}
\eqae
This leads to the usual form of 4 dimensional massless momentum 
\eq
\displaystyle{\not}k_{AB}=\left(\begin{array}{cc}0 & k^\alpha k_{\dot{\alpha}} \\-k_{\dot{\alpha}}k^\alpha & 0\end{array}\right).
\label{4 dimensional momentum}
\eqe 
Note the solutions have definite U(1) helicity. Therefore when the external momenta lie in a 4 dimensional subspace, the connection between 6 dimensions and 4 dimensions little group is now clear: the usual 4 dimensions U(1) helicity group lies in the diagonal subgroup of the 6 dimensions SU(2)$\times$SU(2). One can now relabel the SU(2) indices $a,\dot{a}$ as $\pm$ which represents $\pm\frac{1}{2}$ under the U(1) helicity group, i.e. $\eta_a\rightarrow \eta_\pm,\tilde{\eta}_{\dot{a}}\rightarrow \tilde{\eta}_\pm$.

Another way of viewing this is through the supersymmetric theory. Taking the diagonal subgroup means that in the weight space one projects all states along the diagonal axes
$$\includegraphics{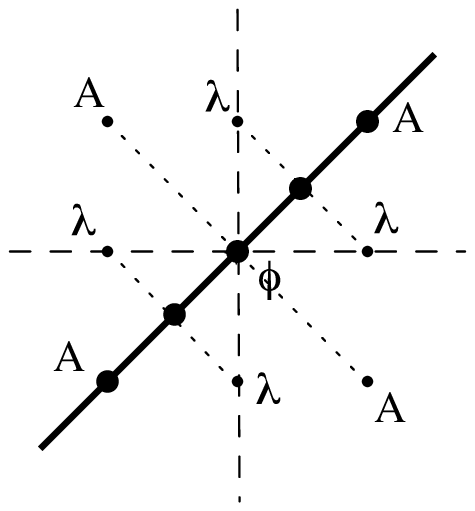}$$
The action of $\eta^a,\tilde{\eta}^{\dot{a}}$ are then projected on this diagonal line and become raising and lowering operators of the U(1) helicity by $\frac{1}{2}$. We then have the following identification:
\eqa
\nonumber &&A^-\sim\eta_-\tilde{\eta}_-,\;\;A^+\sim\eta_+\tilde{\eta}_+,\;\;\phi\sim\eta_+\tilde{\eta}_-,\,\eta_-\tilde{\eta}_+\,(real),\;\eta_+\eta_-,\,\tilde{\eta}_-\tilde{\eta}_+\,(complex)\\
&&\bar{\lambda}\sim\eta_-,\,\tilde{\eta}_-,\;\;\lambda\sim\eta_+,\,\tilde{\eta}_+. 
\eqae 

Now we rewrite all 6 dimensional invariants in terms of 4 dimensional ones

\noindent i) 
\eqa
\nonumber&&\langle i_a|j_{\dot{b}}]=(\lambda_i)^A\,_a(\tilde{\lambda}_j)_{A\dot{b}}=\left(\begin{array}{cc}[ij] &  0 \\ 0 & -\langle ij\rangle\end{array}\right),\;\;[i_{\dot{a}}|j_b\rangle=(\tilde{\lambda}_i)_{A\dot{a}}(\lambda_j)^A\,_b=\left(\begin{array}{cc}-[ij] &  0 \\ 0 & \langle ij\rangle\end{array}\right)\\
\nonumber&&\rightarrow\langle i_-|j_-]=-[i_-|j_-\rangle=[ij];\;\;\;\,-\langle i_+|j_+]=[i_+|j_+\rangle=\langle ij\rangle,\\
\label{brakets}
\eqae 
where as usual $[ij]=(\tilde{\lambda}_i)^{\dot{\alpha}}(\tilde{\lambda}_j)_{\dot{\alpha}}$, $\langle ij\rangle=(\lambda_i)^\alpha(\lambda_j)_\alpha$.

\noindent ii)
\eqa
\nonumber\langle i_aj_bk_cl_d\rangle&\rightarrow&\langle i_+j_+k_-l_-\rangle=-\langle ij\rangle[kl],\,\langle i_+j_-k_+l_-\rangle=+\langle ik\rangle[jl]\cdot\cdot\cdot\\
\nonumber[i_{\dot{a}}j_{\dot{b}}k_{\dot{c}}l_{\dot{d}}]&\rightarrow&[i_+j_+k_-l_-]=-\langle ij\rangle[kl],\,[i_+j_-k_+l_-]=+\langle ik\rangle[jl]\cdot\cdot\cdot\\
\eqae

We demonstrate this with an example, we will derive the known 4 dimensional ($A_1^-,A_2^+,\lambda_3,\bar{\lambda}_4$) amplitude from our 6 dimensional  4-point super amplitude. The 4 dimensional result is  
\eq
4D:\; \mathcal{M}(A_1^-,A_2^+,\lambda_3,\bar{\lambda}_4)=\frac{i\langle14\rangle^3\langle13\rangle}{\langle12\rangle\langle23\rangle\langle34\rangle\langle41\rangle}=i\frac{\langle14\rangle^2[24][23]}{st}
\eqe
We start instead with the 6 dimensional super amplitude:
\eq
6D:\;\mathcal{M}=-i\frac{\delta^4(\sum^4_{n=1}q^M)\delta^4(\sum^4_{n=1} \tilde{q}_M)}{st}.
\eqe
To extract $A_1^-,A_2^+,\lambda_3,\bar{\lambda}_4$ one chooses $d\eta_{1-}d\tilde{\eta}_{1-}d\eta_{2+}d\tilde{\eta}_{2+}d\eta_{3+}d\eta_{4-}$ as integration measure. However, it is obvious there are too many $\tilde{\eta}$ left unintegrated. To introduce additional integration measure and not interfere with the helicity structure, one has only two choices $d\tilde{\eta}_{3+}d\tilde{\eta}_{3-}$ or $d\tilde{\eta}_{4+}d\tilde{\eta}_{4-}$. These two are equivalent up to momentum conservation. Choosing the latter and performing the integration one has 
\eqa
\nonumber\frac{\langle 1_+2_-3_-4_+\rangle[1_+|\displaystyle{\not}p_4|2_-]}{st}&=&\frac{\langle 1_+2_-3_-4_+\rangle[1_+|4_+\rangle\langle4_-|2_-]}{st}\\
&=&\frac{\langle14\rangle^2[23][42]}{st}
\eqae
where we have used the results in (\ref{brakets}).

\section{SU(2) Spinors for 3,4-point calculation} 
\label{equations}
Here we present some of the definitions that are useful in the derivations. For the 3-point amplitude, since the Lorentz invariants $\langle i|j]$ have rank 1, they can be rewritten in terms of SU(2) spinors 
\eqa
\nonumber\langle1_a|2_{\dot{b}}]=u_{1a}\tilde{u}_{2\dot{b}},\;\langle2_a|1_{\dot{b}}]=-u_{2a}\tilde{u}_{1\dot{b}}\\
\nonumber\langle2_a|3_{\dot{b}}]=u_{2a}\tilde{u}_{3\dot{b}},\;\langle1_a|3_{\dot{b}}]=-u_{1a}\tilde{u}_{3\dot{b}}\\
\nonumber\langle3_a|1_{\dot{b}}]=u_{3a}\tilde{u}_{1\dot{b}},\;\langle3_a|2_{\dot{b}}]=-u_{3a}\tilde{u}_{2\dot{b}}\\
\label{not}
\eqae
From momentum conservation,
\eq
\lambda_1\times(p_1+p_2+p_3)=0\rightarrow \langle 1_a|2^{\dot{b}}][2_{\dot{b}}|_A+\langle 1_a|3^{\dot{c}}][3_{\dot{c}}|_A=0\rightarrow \tilde{u}_2^{\dot{c}}[2_{\dot{c}}|=\tilde{u}_3^{\dot{c}}[3_{\dot{c}}|=\tilde{u}_1^{\dot{c}}[1_{\dot{c}}|
\eqe

We will now use these results to demonstrate $w_K\cdot w_P=\frac{1}{\sqrt{-s_{12}}}$.\footnote{This derivation was based on private communication with Donal O'Connell. } As shown in \cite{Cheung:2009dc} one can use the shift degree of freedom to fix 
$$w_K\cdot u_P=w_P\cdot u_K=\tilde{w}_K\cdot \tilde{u}_P=\tilde{w}_P\cdot \tilde{u}_K=0.$$
From definition $u^a_Pw^b_P-u^b_Pw^a_P=\epsilon^{ab}$, one can deduce 
$$(u^a_Pw^b_P-u^b_Pw^a_P)(u_{aK}w_{bK}-u_{bK}w_{aK})=2\rightarrow (u_K\cdot u_P)=\frac{1}{(w_P \cdot w_K)}.$$
Therefore we can instead compute  $u_K\cdot u_P$. We begin by considering the following object:
\eqa
\nonumber \langle \hat{1}_{a}|p_4p_{\hat{2}} |\hat{1}_{\dot{a}}]&=& u_{\hat{1}a}\tilde{u}^{\dot{d}}_{4}[ 4_{\dot{d}}|p_{\hat{2}}|\hat{1}_{\dot{a}}]\\
&=& u_{\hat{1}a}\tilde{u}^{\dot{d}}_{\hat{1}}[\hat{1}_{\dot{d}}|p_{\hat{2}}|\hat{1}_{\dot{a}}]=u_{\hat{1}a}\tilde{u}_{\hat{1}\dot{a}}s_{12}.
\label{claytondown}
\eqae
Where we've used $s_{\hat{1}\hat{2}}=s_{12}$. On the other hand one can also deduce 
\eqa
\nonumber \langle \hat{1}_{a}|p_4p_{\hat{2}} |\hat{1}_{\dot{a}}]&=& u_{1a}\tilde{u}^{\dot{d}}_{P}[ P_{\dot{d}}|p_{\hat{2}}|\hat{1}_{\dot{a}}]= iu_{\hat{1}a}\tilde{u}^{\dot{d}}_{P}[ K_{\dot{d}}|p_{\hat{2}}|\hat{1}_{\dot{a}}]\\
\nonumber &=& iu_{\hat{1}a}(\tilde{u}_{P}\cdot\tilde{u}_{K})u^b_{\hat{2}}\langle \hat{2}_{b}|\hat{1}_{\dot{a}}]=iu_{\hat{1}a}(\tilde{u}_{P}\cdot\tilde{u}_{K})u^b_{K}\langle K_{b}|\hat{1}_{\dot{a}}]\\
&=&  -u_{\hat{1}a}\tilde{u}_{\hat{1}\dot{a}}(\tilde{u}_{P}\cdot\tilde{u}_{K})^2
\label{claytonup}
\eqae
Combining eq.(\ref{claytondown}) and (\ref{claytonup}) and combining the anti-chiral piece we arrive at  
$$(\tilde{u}_{P}\cdot\tilde{u}_{K})=\sqrt{-s_{12}}\rightarrow (w_K\cdot w_P)(\tilde{w}_K\cdot \tilde{w}_P)=-\frac{1}{s_{12}}$$

To express these SU(2) spinors in terms of 4D spinor, one start with $\langle i_a|j_{\dot{b}}]=\left(\begin{array}{cc}0 &  0 \\ 0 & -\langle ij\rangle\end{array}\right)$, $[i_{\dot{a}}|j_b\rangle=\left(\begin{array}{cc}0 &  0 \\ 0 & \langle ij\rangle\end{array}\right)$. Using (\ref{not}) and the definition of $w$, one has 
$$(w_i)_a=\left(\begin{array}{c}1/N_i \\ b_iN_i\end{array}\right)\,(\tilde{w}_i)_{\dot{a}}=\left(\begin{array}{c}1/\tilde{N}_i \\ \tilde{b}_i\tilde{N}_i\end{array}\right)$$
where the definitions of $N_i$ are given in \cite{Cheung:2009dc}, we list them here for convenience:\footnote{With signs appropriate for our convention.}
$$N_2=\frac{\langle23\rangle}{\langle31\rangle}N_1,\;N_3=\frac{\langle23\rangle}{\langle12\rangle}N_1,\tilde{N}_1=-\frac{\langle12\rangle\langle31\rangle}{\langle23\rangle N_1},\tilde{N_2}=-\frac{\langle12\rangle}{N_1},\tilde{N}_3=-\frac{\langle31\rangle}{N_1}$$
The $w$s are defined up to an overall scale $N_1$ and shift parameter $b_i$. Since all the amplitudes derived are invariant under the $b$ shift and $w,\tilde{w}$s come in pairs, the final result is independent of these ambiguities. 

\section{Supersymmetric shift } 
\label{susyshift}
Here we discuss the complex shift that is necessary for the BCFW construction. Taking $1,2$ as the reference lines, we have $\hat{p}_{\hat{1}}=p_1+zq, \,\hat{p}_{\hat{2}}=p_2-zq$ with 
$$q^{AB}=x^a\tilde{x}^{\dot{a}}(\epsilon^{AB}_1)_{a\dot{a}}=x^a\tilde{x}^{\dot{a}}\frac{\lambda^{[A}_{1a}\lambda^{B]}_{2b}}{[1^{\dot{a}}|2_b\rangle}=\frac{|^{[A}x\rangle[\tilde{x}|2^b\rangle\lambda^{B]}_{2b}}{s_{12}}$$
where $|x\rangle=x^a|1_a\rangle$ and $|\tilde{x}]=\tilde{x}^{\dot{a}}|1_{\dot{a}}]$. This shift can be understood as the following shift in the spinor variable of the reference lines
\eqa
\nonumber\lambda^A_{\hat{1}a}&=&\lambda^A_{1a}+zx_a[\tilde{x}|2^b\rangle\lambda^A_{2b}/s_{12}\\
\nonumber\lambda^A_{\hat{2}a}&=&\lambda^A_{2a}+z|^Ax\rangle[\tilde{x}|2_a\rangle/s_{12}\\
\nonumber\tilde{\lambda}_{A\hat{1}\dot{a}}&=&\tilde{\lambda}_{A1\dot{a}}+z\tilde{x}_{\dot{a}}\langle x|2^{\dot{c}}]\tilde{\lambda}_{A2\dot{c}}/s_{12}\\
\nonumber\tilde{\lambda}_{A\hat{2}\dot{a}}&=&\tilde{\lambda}_{A2\dot{a}}+z|_A\tilde{x}]\langle x|2_{\dot{a}}]/s_{12}\\
\label{bshift}
\eqae
To maintain super momentum conservation, one also shifts the Grassmann variables:  
\eqa
\nonumber\eta_{\hat{1}a}=\eta_{1a}+zx_a[\tilde{x}|2^b\rangle\eta_{2b}/s_{12}\\
\nonumber\eta_{\hat{2}a}=\eta_{2a}+z[\tilde{x}|2_a\rangle x^b\eta_{1b}/s_{12}\\
\nonumber\tilde{\eta}_{\hat{1}\dot{a}}=\tilde{\eta}_{1\dot{a}}+z\tilde{x}_{\dot{a}}[2^{\dot{b}}|x\rangle\tilde{\eta}_{2\dot{b}}/s_{12}\\
\nonumber\tilde{\eta}_{\hat{2}\dot{a}}=\tilde{\eta}_{2\dot{a}}+z[2_{\dot{a}}|x\rangle\tilde{x}^{\dot{b}}\tilde{\eta}_{1\dot{b}}/s_{12}\\.
\label{fshift}
\eqae
Therefore we have 
\eqa
\nonumber(\lambda^A_{\hat{1}}\cdot\eta_{\hat{1}})&=&(\lambda^A_{1}\cdot\eta_{1})-z|^Ax\rangle[\tilde{x}|2^b\rangle\eta_{2b}/s_{12}+z[\tilde{x}|2^b\rangle\lambda^A_{2b}x^a\eta_{1a}/s_{12}\\
\nonumber(\lambda^A_{\hat{2}}\cdot\eta_{\hat{2}})&=&(\lambda^A_{2}\cdot\eta_{2})+z\lambda^{a A}_{2}[\tilde{x}|2_a\rangle x^b\eta_{1b}/s_{12}+z|^Ax\rangle[\tilde{x}|2^a\rangle\eta_{2a}/s_{12}\\
\nonumber(\tilde{\lambda}_{\hat{1}A}\cdot\tilde{\eta}_{\hat{1}})&=&(\tilde{\lambda}_{1A}\cdot\tilde{\eta}_{1})-z|_A\tilde{x}][2^{\dot{b}}|x\rangle\tilde{\eta}_{2\dot{b}}/s_{12}+z\langle x|2^{\dot{c}}]\tilde{\lambda}_{2\dot{c}A}\tilde{x}^{\dot{a}}\tilde{\eta}_{1\dot{a}}/s_{12}\\
\nonumber(\tilde{\lambda}_{\hat{2}A}\cdot\tilde{\eta}_{\hat{2}})&=&(\tilde{\lambda}_{2A}\cdot\tilde{\eta}_{2})+\tilde{\lambda}^{\dot{a}}_{2A}z[2_{\dot{a}}|x\rangle\tilde{x}^{\dot{b}}\tilde{\eta}_{1\dot{b}}/s_{12}+z|_A\tilde{x}]\langle x|2^{\dot{a}}]\tilde{\eta}_{2\dot{a}}/s_{12}\\
\label{combo}
\eqae
Note that $(\lambda^A_{\hat{1}}\cdot\eta_{\hat{1}})+(\lambda^A_{\hat{2}}\cdot\eta_{\hat{2}})=(\lambda^A_{1}\cdot\eta_{1})+(\lambda^A_{2}\cdot\eta_{2})$ which is necessary for super momentum conservation.

There is a physical meaning to the parameters $x^a$ and $\tilde{x}^{\dot{a}}$. In the original Yang-Mills calculation, the idea is that even though the shift is defined using the polarization vector of the the 1st leg, the result should not depend on its polarization state\cite{Cheung:2009dc}. $x^a$ and $\tilde{x}^{\dot{a}}$ are arbitrary parameters that parameterize this ambiguity, and the statement that the final result is independent of the polarization state translates into independence of  $x^a,\tilde{x}^{\dot{a}}$. In the supersymmetric case, the first leg may not be a vector. However one still uses the spinors of the first leg to construct polarization vector, which carries an SU(2) little group index. Again the final result should not depend on its state, thus one contracts the SU(2) index of the first spinor to parameterize this dependence, and in the end the final result should again be independent of it.
\section{5-point}
\label{proof}
Here we give some details on the derivation of (\ref{needproof}) and (\ref{lake}): 
\eqa
\nonumber &&-\frac{i}{s_{51}\phi s_{34}}\frac{x^{a} x^{\dot{a}}}{s_{12}}\left[u_{\hat{1}a}\langle2^b|\hat{1}_{\dot{a}}](\tilde{u}_{\hat{1}}\cdot\tilde{\eta}_{\hat{1}}-\tilde{u}_{5}\cdot\tilde{\eta}_5)\right]\left[\tilde{u}_5^{\dot{c}}\langle2_b|5_{\dot{c}}](u_{\hat{1}}\cdot\eta_{\hat{1}}-u_{5}\cdot\eta_5)\right]\\
\nonumber&=&-\frac{ix^{a} x^{\dot{a}}}{s_{51}\phi s_{34}s_{12}}\left[u_{\hat{1}a}\langle2^b|\hat{1}_{\dot{a}}]\tilde{u}_{\hat{1}}\cdot\tilde{\eta}_{\hat{1}}-\langle2^b|\hat{1}_{\dot{a}}]\langle\hat{1}_a|5_{\dot{d}}]\tilde{\eta}^{\dot{d}}_5\right]\left[\langle2_b|5_{\dot{c}}]\langle\hat{1}_d|5^{\dot{c}}]\eta^d_{\hat{1}}-\tilde{u}_5^{\dot{c}}\langle2_b|5_{\dot{c}}]u_{5}\cdot\eta_5\right]\\
\eqae
Now we need to get rid of $u_{\hat{1}}\tilde{u}_{\hat{1}}$. We use:
\eqa
\nonumber u_{\hat{1}a}\tilde{u}_{\hat{1}\dot{a}}&=&u_{\hat{1}b}\tilde{u}_{\hat{1}\dot{a}}\delta_a^b=u_{\hat{1}b}\tilde{u}_{\hat{1}\dot{a}}\langle \hat{1}^b|P_{\dot{b}}](\langle \hat{1}^a|P_{\dot{b}}])^{-1}\\
\nonumber &=&-u_{5b}\tilde{u}_{\hat{1}\dot{a}}\langle 5^b|P_{\dot{b}}]\langle \hat{1}_s|P^{\dot{b}}]/s_{1P}=-\frac{[\hat{1}_{\dot{a}}|\displaystyle{\not}p_5\displaystyle{\not}p_P|\hat{1}_{a}\rangle}{s_{\hat{1}P}}
\eqae
where $p_P$ is an arbitrary null vector. The result: 
\eqa
\nonumber D_1&\sim&\frac{i}{s_{51}\phi s_{34}}\frac{x^{a} x^{\dot{a}}}{s_{12}}\left[\langle2^b|\hat{1}_{\dot{a}}]\frac{[\hat{1}_{\dot{c}}|\displaystyle{\not}p_5\displaystyle{\not}p_P|\hat{1}_a\rangle}{s_{\hat{1}P}}\tilde{\eta}^{\dot{c}}_{\hat{1}}+\langle2^b|\hat{1}_{\dot{a}}]\langle\hat{1}_a|5_{\dot{d}}]\tilde{\eta}^{\dot{d}}_5\right]\left[\langle\hat{1}_d|\displaystyle{\not}p_5|2_b\rangle\eta^d_{\hat{1}}+\langle5_d|\displaystyle{\not}p_1|2_b\rangle\eta^{d}_5\right]\\
\nonumber&=&\frac{i}{s_{51}\phi s_{34}s^2_{12}}\left(\langle x|\displaystyle{\not}p_2\displaystyle{\not}p_5|\hat{1}_{\dot{c}}]\tilde{\eta}^{\dot{c}}_{\hat{1}}+s_{12}\langle x|5_{\dot{d}}]\tilde{\eta}^{\dot{d}}_5\right)\left([\tilde{x}|\displaystyle{\not}p_2\displaystyle{\not}p_5|\hat{1}_d\rangle\eta^d_{\hat{1}}+s_{12}[\tilde{x}|5_d\rangle\eta_5^d\right)
\eqae
where we've chosen $p_P=p_{\hat{2}}$. Similarly for $D_2$
\eqa
\nonumber &&\frac{i}{s_{23}\phi s_{45}}\frac{x^{a} x^{\dot{a}}}{s_{12}}\tilde{u}_{\hat{2}\dot{c}}[\hat{2}^{\dot{c}}|1_a\rangle u_{\hat{2}b}\langle\hat{2}^b|1_{\dot{a}}](\tilde{u}_{\hat{2}}\cdot\tilde{\eta}_{\hat{2}}-\tilde{u}_{3}\cdot\tilde{\eta}_3)(u_{\hat{2}}\cdot\eta_{\hat{2}}-u_{3}\cdot\eta_3)\\
\nonumber&=&\frac{i}{s_{23}\phi s_{45}}\frac{x^{a} x^{\dot{a}}}{s_{12}}\left[\langle3^b|1_{\dot{a}}]\langle3_b|\hat{2}_{\dot{d}}]\tilde{\eta}^{\dot{d}}_{\hat{2}}+\langle\hat{2}^b|1_{\dot{a}}]\langle \hat{2}_b|3_{\dot{d}}]\tilde{\eta}^{\dot{d}}_3\right]\left[[3^{\dot{c}}|1_a\rangle\langle\hat{2}_{d}|3_{\dot{c}}]\eta^d_{\hat{2}}+[2^{\dot{c}}|1_a\rangle\langle3_d|\hat{2}_{\dot{c}}]\eta^d_3\right]\\
\nonumber&=&\frac{i}{s_{23}\phi s_{45}s_{12}}\left[[\tilde{x}|\displaystyle{\not}p_3|\hat{2}_{\dot{d}}]\tilde{\eta}^{\dot{d}}_{\hat{2}}+[\tilde{x}|\displaystyle{\not}p_2|3_{\dot{d}}]\tilde{\eta}^{\dot{d}}_{3}\right]\left[\langle x|\displaystyle{\not}p_3|\hat{2}_b\rangle\eta_{\hat{2}}^b+\langle x|\displaystyle{\not}p_2|3_b\rangle\eta_{3}^b\right]
\eqae



\end{document}